# Exploring Exogenic Sources for the Olivine on Asteroid (4) Vesta

Lucille Le Corre
Planetary Science Institute, 1700 East Fort Lowell, Suite 106, Tucson, AZ 85719, USA.
Email: lecorre@psi.edu

Vishnu Reddy
Planetary Science Institute, 1700 East Fort Lowell, Suite 106, Tucson, AZ 85719, USA.

Juan A. Sanchez
Planetary Science Institute, 1700 East Fort Lowell, Suite 106, Tucson, AZ 85719, USA.

Tasha Dunn
Colby College, Department of Geology, 5800 Mayflower Hill, Waterville, ME 04901

Edward A. Cloutis
Department of Geography, University of Winnipeg, 515 Portage Avenue, Winnipeg, Manitoba, Canada R3B 2E9

Matthew R. M. Izawa
Department of Geography, University of Winnipeg, 515 Portage Avenue, Winnipeg, Manitoba, Canada R3B 2E9

Paul Mann
Department of Geography, University of Winnipeg, 515 Portage Avenue, Winnipeg, Manitoba, Canada R3B 2E9

Andreas Nathues
Max-Planck-Institute for Solar System Research, Göttingen, Germany.

**Proposed Running Head:** Exogenic sources for the olivine on Vesta

**Editorial correspondence to:**
Lucille Le Corre
Planetary Science Institute
1700 E Fort Lowell Rd #106
Tucson, Arizona 85719, USA
lecorre@psi.edu

## Abstract

The detection of olivine on Vesta is interesting because it may provide critical insights into planetary differentiation early in our Solar System's history. Ground-based and Hubble Space Telescope (HST) observations of asteroid (4) Vesta have suggested the presence of olivine on the surface. These observations were reinforced by the discovery of olivine-rich HED meteorites from Vesta in recent years. However, analysis of data from NASA's Dawn spacecraft has shown that this "olivine-bearing unit" is actually impact melt in the ejecta of Oppia crater. The lack of widespread mantle olivine, exposed during the formation of the 19 km deep Rheasilvia basin on Vesta's South Pole, further complicated this picture. Ammannito et al., (2013a) reported the discovery of local scale olivine-rich units in the form of excavated material from the mantle using the Visible and InfraRed spectrometer (VIR) on Dawn. These sites are concentrated in the walls and ejecta of Arruntia (10.5 km in diameter) and Bellicia (41.7 km in diameter) craters located in the northern hemisphere, 350-430 km from Rheasilvia basin's rim. Here we explore alternative sources for the olivine in the northern hemisphere of Vesta by reanalyzing the data from the VIR instrument using laboratory spectral measurements of meteorites. Our rationale for using the published data set was to bypass calibration issues and ensure a consistent data set between the two studies. Our analysis of the VIR data shows that while the interpretation of their spectra as an olivine-rich unit is correct, the nature and origin of that olivine could be more complicated. We suggest that these olivine exposures could also be explained by the delivery of olivine-rich exogenic material. Our rationale for this study is supported by meteoritical evidence in the form of exogenic xenoliths containing significant amount of olivine in some of the HED meteorites from Vesta. Previous laboratory work on HEDs show that potential sources of olivine on Vesta could be different types of olivine-rich meteorites, either primitive achondrites (acapulcoites, lodranites, ureilites), ordinary chondrites (H, L, LL), pallasites, or carbonaceous chondrites (e.g. CV). Based on our spectral band parameters analysis, the lack of correlation between the location of these olivine-rich terrains and possible mantle-excavating events, and supported by observations of HED meteorites, we propose that a probable source for olivine seen in the northern hemisphere are remnants of impactors made of olivine-rich meteorites. Best match suggests these units are HED material mixed with either ordinary chondrites, or with some olivine-dominated meteorites such as R-chondrites.

## 1. Introduction

Olivine is the "Rosetta Stone" of planetary differentiation. Its abundance and chemistry provide important clues into the conditions under which it forms including the composition of the starting material, redox state, temperature, etc. Olivine is also the most abundant mineral formed during planetary differentiation, normally making up the bulk of an object's mantle for inner solar system bodies. However, olivine on asteroids remains elusive with only a handful of asteroids that are dominated by olivine having been identified to date (e.g., Sanchez et al., 2014), and this lack of olivine in the main asteroid belt has been termed the "missing olivine problem".

Using ground-based telescopic observations, several researchers have searched for evidence for the presence of olivine on Vesta without success (Larson and Fink 1975; Feierberg and Drake, 1980). Similarly, McFadden et al. (1977) inferred that olivine was probably very rare or absent from the surface of Vesta using data in the visible wavelength range. The first possible detection of olivine with a significant surface coverage on Vesta was presented in Gaffey (1997). In this paper, the authors interpreted a drop in pyroxene band area ratio (BAR) between 60-120°E longitude (Thomas et al., (1997) coordinate system) in rotationally-resolved spectra of Vesta as a possible indicator of the presence of olivine. This feature, nicknamed the "Leslie formation", was interpreted to be an olivine mantle deposit probably emplaced by ejecta material from a large impact. With the analysis of the Dawn Framing Camera (FC) data, Reddy et al. (2013) associated this "Leslie formation" with the Oppia crater and its ejecta. Le Corre et al. (2013) ruled out the presence of olivine in this crater and concluded that the drop in BAR was due to the presence of impact melt rather than olivine. McSween et al. (2013a) reported that no large olivine deposits were detected within the Rheasilvia basin suggesting that no significant amount of olivine has been excavated by basin-forming impacts in the south polar region. Beck et al. (2013) suggested that the lack of olivine could be explained by the observed dilution of olivine abundance in harzburgitic diogenite exposures.

The scarce detection of olivine on Vesta is consistent with the rarity of olivine-rich diogenites (11 samples of olivine-diogenites and 368 diogenites) in the meteorite collection (Met. Bulletin). Diogenites are generally monomineralic (orthopyroxenite) and contain less than 10% of olivine with minor eucritic fragments (McSween et al., 2010). Olivine diogenites correspond to cumulate orthopyroxene-rich rocks but contain a significant amount of olivine. Among olivine diogenites, Beck and McSween (2010) identified mixtures of olivine and magnesian orthopyroxene as harzburgites, and the most common diogenites made of two distinct orthopyroxenes (Mg-rich and Fe-rich) as breccias of two different lithologies (polymict diogenites). Beck and McSween (2010) proposed new subdivisions among diogenites with dunitic, harzburgitic, and orthopyroxenitic diogenites based on the content of olivine and orthopyroxene: orthopyroxenite has <10% olivine, harzburgite can contain 10-90% olivine, and dunite is characterized by >90% olivine. Only three diogenites composed essentially of pure olivine have been described: a monomineralic dunite clast in NWA 5968 (Bunch et al., 2010) and MIL 03443 (Mittlefehldt, 2008), both brecciated dunites (Beck et al., 2011), and cataclastic dunite NWA 5784 (Bunch et al., 2010). A third achondrite (NWA 2968) with >95 % olivine is also considered a dunite from Vesta (Bunch et al., 2006). Concerning the howardites, Beck et al. (2012) analyzed olivine-rich impact melt in a group of howardites that suggest the possible incorporation of dunitic and harzburgitic lithologies in the howarditic target rock. Two other groups of olivines were found in these howardites, one formed by secondary processes on Vesta (crystallized from the melt), and one possibly from exogenic origin.

Ammannito et al. (2013a) described the identification of olivine-rich units emplaced over howarditic background terrain in Arruntia (39.44°N, 221.59°E) and Bellicia (37.73°N, 197.76°E) craters (for a map with vestan nomenclature, see USGS website[1]). The location of these units comes as a surprise because one would expect olivine to be present around the Rheasilvia basin (e.g. Ruzicka et al., 1997; Jutzi and Asphaug, 2011), which is thought to have breached the crust-mantle boundary (Thomas et al., 1997). Olivine is found in the walls of Arruntia crater, a relatively young crater with well-defined rims and visible ejecta rays; and in the walls of Bellicia crater and nearby secondary craters that have excavated and deposited olivine on Bellicia's ejecta blanket. In addition to crater walls, Ammannito et al. (2013a) described the olivine distribution as being scattered diffusely over a broad area. Bellicia is a larger crater (41.7 km in diameter) than Arruntia (10.5 km in diameter). Bellicia appears older than Arruntia with more subdued crater rims and most of the crater walls and floor covered by regolith. In Dawn Framing Camera (FC) images made with Clementine color ratios, Arruntia exhibits orange/red material in its ejecta (i.e. with redder visible slope than average howarditic terrains) suggesting the presence of impact melt (Le Corre et al. 2013). Both craters are located in a howarditic terrain enriched in eucrite-like material in the FC color ratio map from McSween et al. (2013b), which is in agreement with the results inferred from VIR data (Ammannito et al., 2013a).

[1] http://planetarynames.wr.usgs.gov/Page/VESTA/target

Using color parameters derived from the FC color filters, Thangjam et al. (2014) confirmed the presence of olivine-rich units (> 60 % olivine) in Bellicia and Arruntia craters ranging in diameter from few hundred meters to few kilometers. Arruntia is located ~430 km and Bellicia is located ~350 km from the rim of the Rheasilvia basin. Even though these olivine sites are far from the Rheasilvia basin's floor and walls, Ammannito et al. (2013a) favored a mantle origin for these olivine-rich units because of the large area of the deposits and the lack of detected diogenite-like rock. This seems in contradiction to the generally accepted notion that diogenite from the lower crust would be excavated along with mantle olivine and the fact that all olivine-rich HEDs contain a significant amount of diogenitic material (i.e. orthopyroxenite). Nathues et al. (in press) conducted a global survey of olivine-rich units using FC color data and identified 15 sites in the northern hemisphere corresponding to impact craters with evidence for olivine-rich material on their walls and sometimes ejecta. Based on the geologic settings of these olivine deposits, they conclude a mantle origin for this olivine is unlikely. Additional work focused on the search of olivine-rich units using the VIR data demonstrated the lack of peridotitic mantle inside the Rheasilvia basin (Clenet et al., 2014; Ruesch et al., 2014). In contrast with previous findings, Poulet et al. (in press) proposed that olivine is in fact ubiquitous on Vesta at an abundance of 10-20 % mixed in the howarditic regolith, and presented olivine-bearing howardites containing impact melt as further evidence for this finding.

Understanding the origin of olivine on Vesta has important implications for its formation and evolution. So far several hypotheses have been proposed to explain the lack of significant olivine in the Rheasilvia basin and the detection of olivine in the northern hemisphere. Here we summarize the proposed options:

- If the olivine on Vesta is of endogenic origin, excavated during the formation of the South Pole basins, then it is possible to assess the depth of crust/upper mantle boundary using geomorphological characteristics of the Rheasilvia impact basin. The lack of olivine-rich lithologies (such as harzburgites) in the Rheasilvia basin would suggest that the mantle has not been breached (Clenet et al., 2014) or if it was breached, it is not detectable using the spectral and spatial resolution of the VIR spectrometer as demonstrated by Beck et al. (2013). Combining a VIR mineralogical map of Rheasilvia and Veneneia with results from impact modeling, Clenet et al. (2014) proposed that the crust/mantle boundary (so-called Moho) on Vesta was deeper than originally thought, at a depth of at least 80 km. They suggested that this deep Moho was probably not breached by the two catastrophic impacts (Veneneia and Rheasilvia) at the South Pole. If this were true, then it is unlikely that the deposits of olivine-rich material in the northern hemisphere could come from Rheasilvia ejecta.
- An alternative endogenous source suggested in Ammannito et al. (2013a) to explain olivine-rich units in the northern hemisphere are olivine-rich plutons rising through the upper crust that are being exposed by impacts. This is a preferred scenario presented in Clenet et al. (2014) to explain both the lack of olivine in the material excavated by the Rheasilvia impact basin and the presence of olivine patches in the northern hemisphere. In this scenario, the eucritic crust would be much thicker than inferred from magma ocean crystallization models (e. g., Mandler et al., 2013; Toplis et al., 2013) and the mantle much thinner leading to lower amount of total olivine present on Vesta. The presence of crustal intrusions would imply a more heterogeneous distribution of HED lithologies in the interior of Vesta and has implications for the magma ocean model. Looking at the mineralogy of Vestoids, Mayne et al. (2011) also favored the partial melting option to account for both large-scale homogeneity and small-scale heterogeneity observed on their surfaces. By modeling magmatic intrusions migrating to the surface of Vesta, Wilson and Keil (1996) suggested the possible formation of two types of intrusions: intrusions at shallow depth forming dikes no larger than 10 km vertically and ~1 m in width, and dikes or sills at the base of the crust extending up to 30 km with a thickness up to 3 m. They also demonstrate that large shallow magmatic intrusions should not be common on Vesta, therefore, it is unlikely that a plutonic intrusion can explain multiple hundred-meter patches of olivine scattered over a broad area of ~100 km size. To be consistent with their modeling, we would need to invoke the existence of several small dikes present at shallow depth to explain the distribution of olivine sites. Even if this model (existence of abundant olivine-rich intrusions in the vestan crust) allows the formation of diogenitic intrusions enriched in olivine, it remains unclear why such dikes (or plutons, as discussed in previous papers) won't be exposed within the Rheasilvia basin and its surroundings, which is the largest impact structure on Vesta, and would be observed preferentially in the northern hemisphere. While olivine is spectrally difficult to detect when present at <25 vol.% using the VIR instrument (McSween et al., 2013a), the presence of olivine-rich units with abundances >60 vol.% in the northern hemisphere (Ammannito et al. 2013a; Thangjam et al. 2014) suggests that similar units should be present within the Rheasilvia basin. More recently Palomba et al., (2014) discovered olivine-rich sites in the southern hemisphere with a weaker olivine signature than for Bellicia and Arruntia. This would be in contradiction to the non-detection of olivine by Clenet al. (2014) in the same region.

- Ammannito et al. (2013a) briefly discussed exogenic origin of olivine on Vesta and dismissed the possibility due to the perceived lack of olivine-rich A-type asteroids in the main asteroid belt as potential impactors. However, olivine is present in significant abundance in other asteroids in the main belt as discussed by Sanchez et al. (2014). Ammannito et al. (2013a) also noted that most impactor material should be volatilized upon impact. Evidence for impactor material surviving low velocity collisions on Vesta has been proven by the detection of exogenous carbonaceous chondrite material on the surface (e.g. Reddy et al., 2012a) and in HED meteorites (e.g., Zolensky et al., 1996; Herrin et al., 2011). In addition, the VIR instrument detected a 2.8 μm $OH/H_2O$ feature in the spectra of carbonaceous chondrite-rich dark material (De Sanctis et al., 2013) due to the presence of phyllosilicates and GRaND data showed a hydrogen abundance correlated with the dark material (Prettyman et al., 2012) further constraining its origin. All this evidence points to the probable preservation of impactor material on Vesta for at least some impact events. Clenet et al. (2014) noted that olivine-rich material in the northern hemisphere is excavated by small craters and suggested that this is most likely excavation of pluton material (local olivine-rich layer) as these impacts are too shallow to reach the mantle. They do not consider alternative explanations such as the excavation of pre-existing ejecta blanket material partly made of impactor remnants rich in olivine. In addition, as discussed above, according to modeling by Wilson and Keil (1996), intrusions of magma in the crust would form dikes whereas large-scale shallow magma reservoirs are not likely to form.

In this work we reassess the origin of the olivine-rich unit detected by the VIR instrument onboard the Dawn spacecraft. We analyze the VIR data and compare it to meteoritical and mineral data acquired in the laboratory. Because the VIR team has the best calibrations and data sets, we did not recreate a calibration pipeline and process new data sets. Prior studies did not conduct a detailed mineralogical analysis of the olivine-rich unit using diagnostic spectral band parameters; therefore, we focused on that aspect to explore the origin for the olivine on Vesta.

## 2. Analysis of the VIR spectrum of the olivine unit

Spectra of olivine and mixtures of olivine and orthopyroxene have been studied in detail (Cloutis et al., 1986; King and Ridley 1987; Sunshine and Pieters 1998). The spectrum of pure olivine is dominated by a composite absorption band at 1 µm made of three overlapping individual bands. In contrast, the spectra of HEDs and HED pyroxenes have the characteristic two absorption bands near 1 µm (Band I) and 2 µm (Band II) (Burns, 1970; Burns, 1993). For a given absorption feature, band center corresponds to the wavelength position of the minimum reflectance after continuum removal of the spectrum. Hereafter, the 1-µm absorption band center is called Band I center and similarly the 2-µm absorption band center is called Band II center. The continuum can be computed using a straight line from the shorter to the longer wavelength shoulders of the absorption band, or a polynomial curve fitting the local maxima. Continuum-removal is necessary to allow comparison with laboratory spectral data. For spectra dominated by pyroxene, the Band Area Ratio or BAR is computed by taking the ratio of the area of 2-μm absorption band to the area of 1-μm absorption band. Analyzing BAR can be used to constrain the abundance of olivine or pyroxene in a mixture of olivine+orthopyroxene. As the olivine content increases in a mixture of olivine + pyroxene, the 1-μm absorption band becomes wider and more asymmetric. Gaffey et al. (2002) showed that with increasing olivine content in olivine-pyroxene mixtures, the Band I center shifts to longer wavelength. This would be the case on Vesta where any olivine present would shift the Band I center of pyroxene in diogenites to longer wavelength. Olivine has Band I centers that range from 1.04 μm to 1.094 μm depending on the iron abundance (Sanchez et al., 2014). In contrast, pyroxene has Band I centers between 0.90-0.92 μm for low-Ca orthopyroxene, the dominant mineral in diogenites. The shift in Band I center is a good indicator of the presence of olivine and it will also affect the pyroxene composition derived from the Band I center.

Beck, P. et al. (2011) showed that olivine diogenites cannot be distinguished from the other HEDs in the Band I vs. Band II plot. Their olivine diogenite spectra have only a slightly higher wavelength for the Band I center than typical diogenites and therefore plot on the edge of the field formed by diogenites, leaving them undistinguishable from all the other diogenites. Beck et al. (2013) found that the estimated abundance of olivine on Vesta from putative exposure of a harzburgitic lithology (30 % in the best-case scenario with pristine harzburgite) would not be discernable from orthopyroxenite diogenites using VIR spectrometer data even during the LAMO phase that has the highest spatial resolution (~43 meters/pixel). According to the same authors, olivine could be detectable by the Dawn VIR instrument using the BAR parameter for an exposure of dunitic diogenite (>90 vol.%) over a large area. It is important to note, as explained in Duffard et al. (2004) and Mayne et al. (2011), that the BAR calculation cannot be used to retrieve a precise estimation of the olivine abundance if olivine is mixed with high-calcium pyroxene (HCP), which is a component in eucrites −5.4 to 29% HCP according to Mayne et al. (2010)− and howardites. Mayne et al. (2011) discussed the mineralogy of Vestoids presented in Duffard et al. (2004) and inferred the absence of olivine by using MGM analysis and observation of the Band I center. This is in contradiction with the conclusions drawn by Duffard et al. (2004) using the BAR method from Gaffey et al. (2002).

Ammannito et al. (2013a) inferred the presence of olivine using two primary methods, 1) non-diagnostic curve matching techniques and 2) Band I and Band II centers and the BAR parameter. In their Band I vs. Band II plot, olivine-rich sites are falling off the HEDs' trend line, which is a line where all the HED meteorites plot with each subtype (howardite, eucrite and diogenite) plotting in separate subgroups along this line. For sites identified as olivine-rich, they observed a shift in Band I center to longer wavelength while Band II center remains unchanged relative to background terrain. This is consistent with enrichment in olivine in laboratory spectral data of olivine-orthopyroxene mixtures (Cloutis et al., 1986). Based on the Band II center, they suggest that the olivine-rich sites are primarily associated with howarditic lithology. Ammannito et al. (2013a) derived olivine concentrations in the range of 50-80 vol.% based on spectral matching with continuum-removed spectra of olivine and orthopyroxene mixtures. They attribute the 1.3-µm absorption band (appearing as an inflection) and the weak Band II depth observed in the selected sites to the presence of olivine. This absorption appears also close to the 1.2-µm absorption band observed for some basaltic eucrites (Klima et al., 2008).

*2.1 Curve Matching*

Ammannito et al. (2013a) used curve-matching techniques to compare laboratory spectra of olivine and orthopyroxene mixtures and VIR data to demonstrate the presence of olivine. Characteristics of these mixtures such as the distortion of the shape of the Band I (relative to typical spectrum of orthopyroxene), the occurrence of an inflection near 1.3 µm, and the weaker Band II depth, were used as indicator for the presence of a large amount of olivine. Curve matching is a useful tool to show gross similarities between two spectra but it is also important to note that it is a non-diagnostic tool that cannot provide a unique answer (Gaffey et al., 2002). For example, it does not account for spectral variations due to illumination conditions, grain size and temperature (e.g., Lucey et al., 1998). As argued by Gaffey et al. (2002), a good spectral match might indicate a similar composition but does not preclude a different interpretation. Gaffey et al. (2002) also demonstrated that, to accurately interpret the mineralogy, curve-matching analysis is more critical in spectral regions containing absorption features relative to spectral regions outside absorption features. Therefore we will focus on matching spectral regions containing diagnostic absorption bands.

According to Ammannito et al. (2013a), the olivine-rich areas present a slightly higher-wavelength Band I center, weaker Band II depth, unchanged Band II center and a distorted Band I (between 1.05 and 1.45 μm) relative to the average spectrum of Vesta. Using the curve matching with laboratory mixtures they suggest the presence of olivine at >50 vol.%. We reproduced the curve-matching plot using the digitized data of the olivine-rich unit from their work (Fig. 1) to avoid any discrepancy that could occur due to different method of data processing. We used the olivine-orthopyroxene mixtures with grain size <38 μm (data for larger grain sizes do not match as well the Band I shape and center of the VIR data, and observational data suggest a fine-grained component for the Vesta surface (Hiroi et al., 1995) and removed the continuum with a 2$^{nd}$ order polynomial for each mixture by selecting points at the local maxima for each absorption band. The minor differences observed between our plot and their work are possibly attributed to the different points and methods used for continuum removal.

No single laboratory spectrum of olivine-orthopyroxene mixture can accurately reproduce both the Band I shape and Band I center of the VIR data, as well as the Band II center due to different pyroxene composition. In Fig. 2, we show the range of olivine-orthopyroxene mixtures Ammannito et al. (2013a) suggest as possible matches to the spectrum of olivine unit on Vesta. The shape near 1.3 μm is best reproduced by the mixture containing 80 % olivine but the overall Band I shape with deeper Band I appears to be best fitted with lower olivine content (between 40 and 50 %), although the Band I center does not properly match the data. The Band II depth and shape correspond to ~60-70% olivine while the Band II center is off likely due to a difference in pyroxene composition between the laboratory sample and the Vesta spectrum. In other words, no single olivine-pyroxene mixture is able to accurately match the VIR data of olivine-rich sites on Vesta. This raises the possibility that the spectral match presented by Ammannito et al. (2013a) to show affinity between olivine-pyroxene mixtures with Vesta spectra is not the best analogue for the composition of the olivine-rich unit. To further explore the link between olivine-orthopyroxene mixtures and olivine-rich units on Vesta, we used more robust diagnostic spectral parameters to constrain its mineralogy.

*2.2 Band parameters*

Mineralogical analysis based on diagnostic spectral band parameters is a more robust way to demonstrate affinity between meteorite spectra and spacecraft data. However, in most cases one cannot use the BAR derived from the VIR data to constrain olivine abundance as BAR is affected by the presence of high-Calcium pyroxene (HCP) in eucrites and howardites (e.g. Reddy, 2009). However, as explain below, the effect of HCP on BAR in our case is minimal and can essentially be ignored when trying to constrain olivine abundance.

In the Band I – Band II plot of Ammannito et al. (2013a), the olivine-rich units show their Band I centers shifted to longer wavelength relative to the test area and the HED trendline. They interpreted this shift in Band I center to be the result of olivine with an abundance of ~60 vol.% in these areas. Looking at the Band II center in the same plot, they infer that these terrains are howarditic in composition. We retrieved the band centers of the olivine-orthopyroxene mixtures used in the previous section and compared them with the VIR data (Table 1). For deriving the band centers from the VIR data, we used third order polynomial fit for each absorption band. We obtained 0.954 ± 0.002 $\mu$m for the Band I center and 1.973 ± 0.007 $\mu$m for the Band II center. The olivine-rich unit Band II center is in the range of Band II centers of eucrites (1.97-2.05 $\mu$m), which are higher than Band II centers of diogenites because of a higher content in Fe-rich pyroxene in eucrites (Burbine and Buchanan, 2010). Therefore the unit might be consistent with a eucrite-rich howardite composition. Using the same Matlab-based code as in Le Corre et al. (2013), we found that the corresponding Band Area Ratio is 0.460± 0.025.

Comparing the Band I centers from the olivine-orthopyroxene mixtures, the Band I center for the olivine-rich site corresponds to 70-80 % of olivine (Table 1). In addition, the BAR value of the olivine-rich site suggests a range between 70-80 vol.% olivine. The Band I center and BAR give olivine abundance between 70-80 vol.%, which is consistent with olivine content in the range of 50-80% obtained by Ammannito et al. (2013a). However, as discussed in the previous section, none of the olivine-pyroxene mixtures accurately reproduce the shape of the VIR spectrum from the olivine-rich unit. In addition, it is important to note that the chemistry of the pyroxene(s) in the regolith can skew the olivine abundance estimate retrieved using Band I center. The abundance of HED component (HCP abundance) in the assemblage can affect the BAR of the olivine-rich unit and subsequently the olivine abundance derived from it. However, the influence of HCP on the BAR is rather limited in our case. We know that the abundance of HCP in eucrites ranges from 5-29 vol.% (Mayne et al., 2010). Delaney et al. (1983) define howardites as those containing 10-90 vol.% eucritic material. If we assume that the background howardite has a maximum of 90 vol.% eucritic material then the abundance of HCP in the background material is 26.1 vol.%. The abundance of olivine is estimated to be ~70% if we use olivine-OPX mixture curve match like Ammanito et al. (2013a). So the abundance of background howarditic material based on the olivine-OPX curve match is ~30 vol.%. If we assume that 90 % of this background howarditic material is eucrite (27 vol. % of the total), then the abundance of HCP in the total mixture is 7.83 vol. % compared to 70 vol.% olivine. In other words, the abundance HCP in this mixture is nine times lower than olivine and can essentially be ignored as its influence on BAR is very low. If we assume that the VIR spectrum represents a near pure phase of exogenic meteorite rich in olivine (maybe OC), then the influence of HCP is even lower than the previous option. BAR might be the least affected band parameter to estimate the abundance of olivine/olivine-rich material.

Our analysis of the VIR data published by Ammannito et al. (2013a) shows that while the interpretation of their spectra as an olivine-rich unit might be correct, the nature and origin of that olivine might be more complicated. In the next section we summarize our knowledge of olivine sources in the main asteroid belt and which of these sources could be potential contributors to Vesta assuming an exogenic origin. Our rationale for suggesting an exogenic source is based on the lack of coherent hypothesis that explains the origin of olivine on Vesta by previous authors as summarized in Section 1.

## 3. Alternative Sources of Olivine

### *3.1 Olivine-rich Asteroids and Meteorites*

While olivine is one of the most ubiquitous minerals found in asteroids and meteorites, olivine-dominated asteroids and meteorites are relatively rare. Olivine-dominated asteroids are typically classified as S(I)-type under the designation introduced by Gaffey et al. (1993), or alternatively are classified as A-types by most of the taxonomic systems (e.g., DeMeo et al., 2009). This population of asteroids spans a wide range of thermal processing, from the most primitive bodies that experienced no significant heating, to those that underwent melting and differentiation. Primitive olivine-dominated asteroids are thought to have formed through nebular processes from accretion of grains in a highly oxidized region, undergoing little post-accretionary heating. This type of object is represented in the terrestrial meteorite collection by the highly oxidized and brecciated R-chondrites, which are characterized by having an olivine chemistry of $Fa_{37-40}$, with traces of low-Ca pyroxene and volumetrically minor clinopyroxene component (Schulze et al., 1994).

On the other extreme, olivine could also be a product of planetary differentiation. Brachinites are partial melt residues that are left behind after the extraction of a basaltic melt. These meteorites are a group of primitive achondrites containing 80-90 % olivine ($Fa_{25-40}$), 5-10 % clinopyroxene and some traces of low-Ca pyroxene (Mittlefehldt et al., 2003). However, their origin is still debated between oxidized and recrystallized chondritic materials, partial melt residues or igneous cumulates (Weisberg et al., 2006). Two other types of primitive achondrites, Acapulcoites and Lodranites, represent residues from partial melting as well, and contain close to chondritic abundances of olivine (Weisberg et al., 2006). If the parent body experienced complete or near-complete melting, on the other hand, then olivine would crystallize from the magma leading to the formation of an olivine mantle. The catastrophic disruption of a differentiated body is thought to produce fragments from such olivine-rich planetesimal mantles and form olivine-dominated asteroids. Fragments from the boundary between olivine mantles and the metallic cores have long been thought to be represented by pallasite meteorites, though recent results indicate that most pallasites may actually result from the collisional disruption and reassembly of differentiated asteroids (Yang et al., 2010). Pallasites are composed of Fe, Ni metal and olivine ($Fa_{10-20}$) in roughly equal amounts with troilite as a minor phase (Mittlefehldt et al., 1998). Ureilites, a rare type of olivine-pyroxene achondrites, have also been proposed as possible analogues for olivine-dominated asteroids (e.g. Goodrich et al., 2014). Three types of ureilites exist: olivine-pigeonite, olivine-orthopyroxene and polymict ureilites (Weisberg et al., 2006). They are composed of Mg-rich olivine ($Fa_{6\text{-}26}$) and minor clinopyroxene, which are embedded in a dark matrix of carbonaceous material (graphite), metal and sulphides (Mittlefehldt et al., 1998).

Olivine-dominated asteroids can be found from the main belt to the near-Earth asteroid population (Sunshine et al., 2007; Reddy et al., 2011; Sanchez et al., 2014). Sunshine et al. (2007) used the Modified Gaussian Model (MGM) to derive the olivine composition of some of these objects. They found that most of the studied objects are Mg-rich with compositions similar to that of pallasites. However, two of the asteroids in their study were found to be more ferroan with compositions consistent with R chondrites. Sanchez et al. (2014) distinguished two classes of olivine-dominated asteroids, one class that they called monomineralic-olivine asteroids, whose spectra exhibit the 1 μm feature and no detectable 2 μm feature, and another class referred to as olivine-rich asteroids, whose spectra exhibit the 1 μm feature and a weak 2 μm feature. They derived the olivine chemistry and mineral abundances using a new set of spectral calibrations from Reddy et al. (2011). For the monomineralic-olivine asteroids they found olivine chemistries in the range of $Fa_{30\text{-}51}$, consistent with both brachinites and R chondrites. Olivine-rich asteroids, on the other hand, showed spectral characteristics similar to those of R chondrites, with olivine abundances ranging from 68 to 78% (Bischoff et al., 2011). Interestingly, several of the objects studied by Sanchez et al. (2014) are located at ~ 2.8 AU from the Sun, near the 5:2 mean motion resonance with Jupiter. Because this region is capable of transporting asteroids to the near-Earth space (Gladman et al., 1997; Nesvorný et al., 2009), it is plausible to think that one of these objects could have impacted Vesta while traversing through the inner part of the main belt.

Asteroids with an ordinary chondrite-like composition also contain a significant fraction of olivine. These objects are classified as S(IV) by Gaffey et al. (1993), and are associated to the S-complex and Q-types (DeMeo et al., 2009). Ordinary chondrites are divided into three subgroups (H, L, LL) based on their Fe content and the ratio of metallic Fe ($Fe^0$) to oxidized Fe (FeO) (Weisberg et al., 2006). X-ray diffraction (XRD) measurements of ordinary chondrite samples show a different range of olivine abundances for each type, with values of 29-40 wt% for H, 38-45 wt% for L, and 47-57 wt% for LL chondrites (Dunn et al., 2010a). A parent body or asteroid family has been associated with each of these subgroups. For example, asteroid (6) Hebe, located at ~ 2.43 AU from the Sun, has been recognized as one of the parent bodies for H chondrites (Gaffey and Gilbert, 1998). In contrast, L chondrites seem to have originated farther away in the main belt and have been linked to the Gefion family, located at ~ 2.8 AU (near the 5:2 resonance) from the Sun (Nesvorný et al., 2009). In the case of LL chondrites, the most likely source region is the Flora family, located in the inner part of the main belt at ~ 2.3 AU from the Sun (Vernazza et al., 2008, de León et al. 2010, Dunn et al., 2013).

Cosmic ray exposure ages of H chondrites show that their parent body experienced multiple collisional events (Graf and Marti, 1995). Therefore, if 6 Hebe is the parent body of these meteorites, fragments of this asteroid should be distributed between the $\nu_6$ secular resonance (at ~ 2.05 AU) and the 3:1 mean motion resonance (at ~ 2.5 AU). Dynamical models show that both of these resonances are capable of moving objects from their respective locations to the near-Earth space (e.g., Morbidelli and Nesvorny, 1999; Morbidelli et al., 2002). Thus, fragments of (6) Hebe moving from the 3:1 resonance could eventually collide with Vesta in their route to the inner regions of the Solar System. It is important to note that no dynamical family is associated with (6) Hebe although several NEAs with compositions similar to H chondrites have been observed (Dunn et al., 2013; Thomas et al., 2014; Kelley et al., 2014). Something similar could happen with L chondrite fragments moving from the 5:2 mean motion resonance with Jupiter (see discussion above). While the Flora family is located close to the Vesta family, the likelihood of an LL chondrite colliding with Vesta seems to be lower. The Flora family is distributed from ~ 2.17 to 2.33 AU, with an $e_p$ in the range of 0.109-0.168, and $i_p$ in the range of 2.4-6.9° (Zappalá, 1995), while Vesta is located at ~2.36 AU, with an $i_p$ of 6.35°. Therefore, the proximity of the Flora family to the $\nu_6$ secular resonance would favor the delivery of objects to inner regions of the Solar System and not to orbits with a > 2.33 AU, making a collision between Vesta and one of these objects less likely.

Cloutis et al. (in press) provide a complete list the meteorite types containing olivine+metal as the dominant component (primitives chondrites, differentiated groups and impact-generated meteorites), and a list of meteorites types with olivine as the dominant mafic silicate phase and with low metal abundance (angrites, ureilites, R chondrites, olivine diogenites, brachinites, chassignite Mars meteorites, CR, CO, CV, CK and CH carbonaceous chondrites). Among these olivine-bearing meteorite types, Cloutis et al. (in press) point out the ones showing a reflectance spectra dominated by the olivine absorption features: CH, CK, CR, CV chondites, ordinary chondrites, R chondrites, angrites, brachinites, chassignites, lodranites, olivine diogenites, pallasites and ureilites.

Among the chondrites, the CV (Vigarano-like) meteorites contain significant amount of olivine. All the CV chondrites are divided in three subgroups according to their petrology and show varying degrees of late-stage alteration that may correspond to different lithologies found on the CV parent body (Weisberg et al., 2006). CV chondrules are rich in olivine and it also is the dominant phase in CV chondrite matrices (Brearley and Jones, 1998). Olivine in CV matrices has varying compositional ranges depending on the meteorite and its origin between solar nebula condensates and secondary processes (aqueous alteration and metamorphism) is still debated (Brearley and Jones, 1998). CV chondrites are typically made of ~75-85 vol.% of olivine (Howard et al., 2009). For example, the Allende meteorite is made of ~84 vol.% of olivine (Bland et al., 2004).

### *3.2 Olivine-rich exogenous material in the HEDs*

HED meteorite analogues for the olivine-bearing units seen by VIR could correspond to polymict breccias composed of lithified fragments, clasts and matrix representing both the stratigraphy of the target body and the relative abundance of types of impactor material (Bischoff et al., 2006). Warren et al. (2009) further classified specific types of howardites that contain high siderophile abundance, foreign fragments and impact melt particles as regolithic howardites. These characteristics are generally in accordance with noble gas enrichments indicative of their time spent in the upper one meter of the vestan regolith (Rao et al., 1991; McSween et al., 2010). The most common xenolithic fragments found in polymict breccias from Vesta are carbonaceous chondrites (Chou, 1976). Lorenz et al. (2007), McSween et al. (2010), and Janots et al. (2012) summarized past work on analysis of exogenous clasts in HEDs, showing that no less than five different carbonaceous chondrite clasts (CV, CM, CR, CI and CK) have been recognized. However, CM chondrites appear dominant (~80% of the carbonaceous clasts) with CR being the next most common carbonaceous clast (~20 %) (Zolensky, 1996; Lorenz, et al., 2007; Janots et al., 2012). Using estimated Ni contents for pristine basaltic or cumulate eucrites, and pristine diogenites, McSween et al. (2010) inferred that howardites and polymict eucrites should contain between 0 and 4% chondritic debris from CM impactors. In some howardites such as PRA 04401, carbonaceous chondrite content is as high as ~60 vol.% (Herrin et al., 2011).

Lorenz et al. (2007) also identified other type of clasts (ordinary chondrite, enstatite chondrite, aubrite, ureilite and mesosiderite) in howardites and polymict eucrites; but they observed that fragments of ordinary chondrites and differentiated meteorites only account for ~1% of meteorite xenoliths in HED breccias. Lorenz et al. (2007) proposed the idea that this lower abundance of metallic meteorites such as mesosiderites and ordinary chondrite could be explained by the fact that these types of impactors originated mostly from high-velocity break-up of their parent body and most would be vaporized upon impacting Vesta unless, they were delivered by a grazing impact. Lorenz et al. (2007) based their identification of ordinary chondrite fragments on the presence of metal-silicate aggregates, metal-troilite inclusions and chondrule-like objects. Lorenz et al. (2007) found one ureilite fragment in Dhofar 018 howardite recognized by its $Cr_2O_3$ rich silicates, and interpreted to be likely produced by impact melting of an ureilite-like impactor (or made of ureilite low-melting components) to explain the higher proportion of pyroxene of the fragment instead of olivine dominated like typical ureilites. In the same howardite, Lorenz et al. (2007) described another fragment likely made of shock-melted aubrite showing no chondritic texture and low modal metal content. Finally, they also suggested that a fragment of metal-silicate aggregates could correspond to a primitive achondrite (acapulcoite or lodranite), and suggested that several xenoliths could come from mesosiderite meteorites or HED silicates mixed with iron meteorites.

Wee et al. (2010) analyzed 13 howardites and 4 polymict eucrites to catalog exogenic impactor materials in HED breccia. They used PGEs (platinum group elements) as an indicator of exogenous contamination in the HED breccia, knowing that eucrites and diogenites have a very low PGE content. Wee et al. (2010) found the PGE abundance and ratios are not the same for all howardites and polymict eucrites. Even though they concluded that it is difficult to identify the exact type of impactor, PGE measurements indicate the possible contamination of HED breccias with a variety of chondritic materials including CI, CK, CM, EH, EL, H, L and LL chondrites. A peculiar howardite, NWA 3197, is a 324-g polymict breccia containing clasts and crystal debris from several types of eucrites and diogenites (Met. Bulletin #102). More interestingly, NWA 3197 exhibits large clasts of recrystallized and shocked H chondrite material with various degree of shock alteration. Composition of the olivine in chondritic clasts is $Fa_{15.3-18.9}$. The ordinary chondrite component seems to account for a significant portion of the meteorite but no detailed petrologic description has been published to date. This meteorite, although interesting as an analogue material for our study, appears to bear a lot of rust and would not be suitable for near-IR spectroscopy measurements in order to compare with VIR data.

A group of howardites found at the Pecora Escarpment Icefield (PCA) in 2002 contains olivine formed from impact melting of dunite- and harzburgite-rich target rocks (with $Fa_{50}$) and from exogenic contamination (with $Fa_{15}$) (Beck et al., 2012). Given the nature of the target material, Beck et al., (2012) suggested that olivine-rich lithologies have been exposed at the surface of Vesta. They also argued that no other olivine composition with $Fo_{85}$ was found in 13 of the 16 olivine-bearing diogenites in the Antarctic meteorite collection, and that the Fe/Mn of this olivine is lower than for diogenitic olivine, pointing to an exogenous origin for this material. Given the abundance and composition of the metal in this sample, they suggest the contamination of the howardites with a metal-rich meteorite such as a pallasite (Beck et al., 2012). Lunning et al. (2014) analyzed olivine grains present in regolithic howardites to characterize Vesta's magmatic history assuming a mantle origin for these grains ($Fa_{8-20}$). However, it is interesting to note that this range of olivine chemistries is close to what is found for H-chondrites (Dunn et al., 2010b), and that excavation of these Mg-rich olivine grains by the Rheasilvia impact event is unlikely given the evidence that the vestan mantle has not been breached (Clenet et al., 2014).

Significant contamination of HEDs with exogenic ordinary chondrite material has been recently brought to light by Janots et al. (2012), who analyzed the howardite Jiddat al Harasis (JaH) 556. With its oxygen isotope composition anomalous for an HED and its enriched siderophile element contents, this points to a contamination by material foreign to Vesta (Janots et al., 2012). According to Janots et al. (2012), the oxygen isotope composition of the bulk rock corresponds to a mixture of HED material and exogenous material with a high $\Delta^{17}O$. JaH 556 is a polymict breccia with impact melt, highly shocked clasts, and remnants of chondrules with a core olivine composition ($Fa_{20}$) similar to an H chondrite. All this evidence leads the authors to conclude that this breccia is made of 20% H chondrite material. Because some of the original chondritic material is dissolved in the impact melt, the abundance of H chondrite material derived from petrographic observation is lower than the one derived from siderophile elements analysis (Janots et al., 2012). They note that meteorites resembling HEDs but with an anomalous oxygen isotopic composition could have been contaminated by non-HED exogenic material. Greenwood et al. (2012) suggested that Dhofar 007, a polymict breccia, was formed by impact mixing based on its high siderophile element content and very heterogeneous oxygen isotope composition. Greenwood et al. (2012) also note that the anomalous oxygen isotope composition for some other previously described anomalous HEDs could be the result of contamination by impactor material and impact mixing processes rather than non-Vesta parent bodies.

Table 2 presents the olivine content of different type of olivine-bearing meteorites along with a summary of their presence in the HED meteorites from the literature. Given the types of exogenous material observed in HEDs and their respective olivine content, possible olivine-rich meteorites that could be analogue candidates for the olivine-rich unit in Vesta's northern hemisphere are primitive achondrites (acapulcoites, brachinites, lodranites, ureilites), R chondrites, ordinary chondrites (H, L, LL), and carbonaceous chondrites (e.g. CV). Even though no brachinite or R chondrite clasts have been found in HED meteorites, we analyzed the spectra of these meteorites to compare with the olivine-rich sites since they have high abundance in olivine and have been detected among the A-type asteroids (Sanchez et al., 2014).

### 3.3 Morphologic evidence for an exogenic origin

The olivine detections presented in Ammannito et al. (2013a) are linked to Bellicia and Arruntia impact craters, particularly in the crater walls and ejecta blankets. FC and VIR image mosaics of both craters are available in Ammannito et al. (2013a) and Thangjam et al. (2014). These two craters may be excavating material from a pre-existing heterogeneous sub-surface layer enriched in olivine (such as an older ejecta blanket covered by howarditic regolith). Arruntia has a morphology corresponding to a younger crater relative to Bellicia, so the olivine detected in the Arruntia crater could correspond to material excavated from the ejecta blanket of the Bellicia crater. However, these two craters may be too far apart for this scenario to happen.

If we consider the case in which the impactor that formed the Rheasilvia basin was an asteroid made of olivine-rich material, Bellicia or Arruntia, being younger craters than Rheasilvia, could have excavated material from the Rheasilvia ejecta blanket, exposing a mixture of vestan regolith and ordinary chondrite impactor remnants. However, Bellicia and Arruntia are located at higher latitude than the northernmost part of the Rheasilvia ejecta lobe. Rheasilvia ejecta material corresponds to the swath of bright albedo and deeper Band I depth material visible partially in the first FC maps from the approach phase in Reddy et al. (2012c), and almost fully in the HAMO phase map presented in Le Corre et al. (2013). Therefore, Rheasilvia ejecta blanket cannot explain the olivine-rich units exposed in the crater walls of Bellicia as observed by Ammannito et al. (2013a).

Given the fact that there is no widespread detection of olivine-rich material near the rim or on the walls of the Rheasilvia basin (e.g., McSween et al., 2013a), this would suggest that an olivine-rich subsurface unit could have been deposited during the formation of an impact crater located close to Arruntia and Bellicia, such as 91 km crater Albana (76.61°N, 200.69°E). In this case, the exogenous material enriched in olivine would have originated during the formation of Albana crater. The ejecta, consisting of a mixture of impactor remnants, impact melt and howarditic material was subsequently covered by impact gardening and then re-excavated by the Bellicia and Arruntia impact events. In this case, one impactor would be necessary to contaminate the area with olivine. To preserve significant material from the impactor, the impact has to occur either with a low velocity or with an oblique angle (Piezzaro and Melosh, 2000; Jutzi and Asphaug, 2011). A similar case has been proposed by Reddy et al. (2012a) in which the contamination of the surface of Vesta with carbonaceous chondrite material was due to a low-velocity impactor forming the Veneneia impact basin. Using the work of O'Brien and Sykes (2011), they show it is possible that some impacts can occur at low velocity at Vesta to deliver exogenous material to the surface and that the preserved impactor material would be diluted in a smaller volume of target material relative to a higher velocity impact. Delivery of exogenous olivine via the same scenario could explain the enriched olivine units found in and around impact craters Arruntia and Bellicia. The differences in the olivine-rich exposures between Arruntia and Bellicia presented in Thangjam et al. (2014) could be due to the heterogeneous distribution and thickness of the ejecta blanket of Albana; or due to the more eroded appearance of the Bellicia crater (older crater) with downslope movements of regolith that could cover olivine-rich exposures. Excavation of olivine-rich material by an old impact basin in the northern hemisphere was also proposed by Thangjam et al. (2014) but in the context of the magma ocean model assuming that the olivine originated from the vestan mantle. They also considered the option of exogenic olivine, but only regarding A-type asteroid impactors, and do not favor the latter explanation due to the rarity of these types of asteroids, similar to what is argued by Ammannito et al. (2013a). A companion paper to this work by Nathues et al. (in press) presents a detailed analysis of the geological settings of the olivine detections on Vesta using Framing Camera data that also supports the exogenous hypothesis for several of the olivine-rich sites with a possible delivery of olivine by the Albana impact event.

## 4. Testing Exogenic Olivine Hypothesis

### *4.1 Curve Matching*

Like our previous attempt with the laboratory mixtures of olivine+orthopyroxene, we started off with a rudimentary curve matching technique using a spectral library of meteorite candidates discussed in section 3.2. The following types of meteorite types were selected for comparison based on their olivine content and possible identification as foreign materials in HEDs: ordinary chondrite, acapulcoite, lodranite, ureilite and carbonaceous chondrite. R chondrite and brachinites, although not detected in HEDs, have been identified in A-type asteroids (Sanchez et al., 2014) and are also selected for our analysis. For each meteorite type, we extracted from the RELAB database all the meteorite spectra acquired on samples with various grain sizes, excluding spectra obtained for slice samples. We then attempted to find the best match to the VIR data after removing the continuum using 2nd or 3rd order polynomials. Despite its limitations, curve-matching results give us a hint on the types of meteorites that can best reproduce the shape of the olivine-rich unit spectrum.

We first compared the laboratory spectra of primitive achondrites with the VIR data. RELAB spectra of Acapulcoites and Lodranites are presented in Fig. 3a. We can notice the effect of changing grain size on the pyroxene absorption bands for the two samples of Lodran at <125 μm and <1000 μm. Among the six meteorites, we visually inspected the continuum-removed spectra against the VIR data and show the best matches (Lodran, Acapulco, and ALHA81261) in Fig. 3b. The Band I shape is close to the VIR data but the asymmetric shape due to the presence of olivine is not pronounced enough in these meteorites. The Band I centers of these spectra are close to the Band I of the olivine-rich unit but the Band II is shifted relative to the VIR data, likely due to a different pyroxene composition or to temperature effects. Fig. 4a shows all the ureilite spectra from the RELAB database. All of them have pyroxene band depths much shallower than the VIR data. The weak absorption bands are attributed to the presence of dispersed graphite and related carbonaceous phases, in addition to metal and maybe shocked material (Cloutis et al., 2010). The most interesting set of spectra are presented in Fig. 4b. These exhibit similar Band I and Band II centers to the olivine-rich unit; however the asymmetry of the Band I does not match the one seen in the VIR data.

RELAB spectra of H, L and LL ordinary chondrites (Dunn et al., 2010b) are shown in Fig. 5a, b and c, respectively. Ordinary chondrites exhibit a Band I asymmetry similar to the olivine-rich unit on Vesta, with a more pronounced asymmetry as the olivine content increases from H, L to LL subtype. Three selections of the best match for each sub-type are presented in Fig. 5d, e and f. The meteorite spectra with the best fit to the VIR data are H5, H6, L6, LL4 and LL5 ordinary chondrites. The shape of the Band I of the VIR data is best reproduced by H6 Canon City (Fig. 5d), the only ordinary chondrite with matching Band I depth, but the Band I center seems to be closer to that of the L6 meteorites (Fig. 5e). As for the Band II center and depth, either sub-type of ordinary chondrites matches the VIR data reasonably well. Based on curve matching, near-pure ordinary chondrite material such as H type could be a good analogue for the olivine-rich unit on Vesta.

RELAB spectra of olivine-dominated meteorites are presented in Fig. 6 (R-chondrites), Fig. 7 (pallasites), Fig. 8 (CV chondrite), Fig. 9 (brachinite). Spectra of R-chondrites and the one example of brachinite have very little to no 2-μm pyroxene absorption and are dominated mostly by the absorption feature of olivine at 1 μm. Spectra of pallasites exhibit only the olivine absorption features near 1 μm, whereas the example of CV chondrite's spectrum exhibits Band I and Band II with very shallow band depths. CV chondrites are very rich in olivine but their reflectance is low: ranging from ~5% to 13% at 0.56 μm and ~5% to 15% at 0.7 μm (Cloutis et al., 2012). CV chondrites have lower reflectance than the olivine-rich sites found near the Bellicia and Arruntia craters. According the FC color data shown in Thangjam et al. (2014) the reflectance of the olivine-rich sites is comprised between 0.23-0.29 at 0.555 μm and between 0.25-0.3 at 0.750 μm. Therefore, if present, CV chondrite material should be mixed with the target material (howardite) in a small quantity as to not darken the regolith too much. Addition of a small quantity of opaque minerals can dramatically reduce the albedo of the vestan regolith (e.g., Reddy et al., 2012a). It seems in contradiction with the strong spectral signature of olivine observed using VIR, which implies that a lot of olivine is exposed and would require a lot of dark CV chondrite material to be present. Olivine-rich impact melt derived from CV material such as the one described in Bullock et al. (2013) could be a potential source of olivine as well but little is known about the occurrence of such material among CV chondrites and about their reflectance. For the other olivine-dominated meteorite candidates (brachinite, R-chondrite and pallasite), which show a strong olivine spectral signature, mixtures with HED meteorites might be more appropriate for curve-matching with the VIR data.

We computed areal (synthetic) mixtures of olivine-dominated meteorites as well as ordinary chondrites with an HED meteorite (eucrite NWA 7465) in an attempt to obtain a better fit to the VIR data (Fig. 10). This is based on the fact that any exogenic material on Vesta would be mixed with the howarditic regolith, as we observed in the case of remnant carbonaceous chondrite impactors (Reddy et al. 2012a). We chose eucrites as our base for the mixtures with ordinary chondrites because the background howardite terrain of the olivine-rich unit has a Band II center within the range of eucrites, and also because we found a better match of the right side of the Band I using basaltic eucrites rather than howardites to fit the VIR data. Eucrites generally have deeper band depths than the howardites we tested in areal mixtures. Mixtures of eucrite with ordinary chondrites are shown in Fig. 10a with the best match to the VIR data being the mixtures containing Tuxtuac (LL5). For mixtures made with Canon City (H6) and Kyushu (L6), the addition of eucrite still leaves the curve slightly higher than the VIR data on longer wavelength side of the Band I. For the mixture with Canon City, Band I and II depths are a bit too deep compared to the VIR spectrum. Shock darkening and the presence of impact melt can strongly affect the absorption band depth without changing the composition of the parent material (Gaffey, 1976; Adams et al., 1979; Reddy et al., 2014). Therefore, this could indicate the presence of shocked/melted regolith (with shallower Band I and Band II depths) mixed with the olivine-rich material. Alternatively, such a band depth discrepancy could be accounted for by invoking grain size differences between laboratory samples and the actual regolith. Other mixtures with olivine-dominated meteorites (pallasite, R-chondrite and brachinite) are shown in Fig. 10b and also exhibit good curve matching with the VIR data. For all the mixtures in Fig. 10a and b, the Band II depth is always deeper relative to the Band II depth for the olivine-rich unit. This could be due to the choice of this specific unaltered eucrite meteorite for computing the mixtures. Minor differences in band depth between the two data sets could also be explained by the lack of photometric correction in the VIR data, i.e. the illumination angles may be different than those of the laboratory data acquired at incident angle of 30° and an emission angle of 0°. It is also well known that observation geometry affects band depths, specifically the relative intensity of Band I and Band II depths decrease with increasing phase angle (e.g., Beck, P. et al., 2011; Reddy et al., 2012b;

Sanchez et al., 2012). Therefore, it is not surprising to observe some discrepancies in band depth, and one has to be cautious when interpreting band depth in spacecraft data that has not been corrected for photometric effects.

*4.2 Band parameter analysis*

A more robust way to establish compositional affinity of the olivine-rich unit is to carry out mineralogical analysis based on diagnostic spectral band parameters. In this section we use band parameters of the VIR data and compare them with band parameters of the possible olivine-rich meteorites we selected (Table 3). For the ordinary chondrites, we used band parameters from the work of Dunn et al. (2010b). Band I center is a close match to six L chondrites (such as L4 Rupota and L5 Malakal) as well as one H chondrite (Canon City), whereas the best match for Band II is found with four LL chondrites (e.g. LL4 Hamlet and LL5 Olivenza). The ordinary chondrite BAR values closest to the BAR for the olivine-rich site are found in the LL chondrites (LL4 Witsand Farm), but some L chondrites, such as L6 APth, have also a similar BAR value (~0.51).

First we plotted the band parameters of the RELAB laboratory spectra (Table 3 and Table 4) in the Band I center vs. Band II center plot in Fig. 11a. HED meteorites from Table 4 are located on an easily visible trendline. Ordinary chondrites subtypes form another trendline with a bit more scatter in the data points, but each are still easily distinguished from one another. Primitive achondrites are scattered in the vicinity of the HED points, whereas the olivine-dominate meteorites are located at the top of the plot near the highest Band I center values. The olivine-rich unit plots in the L chondrite region (Fig. 11b), above the HED trendline, since the Band I center has a higher value due to the presence of olivine. It is also located quite close to some ureilites.

We also plotted the spectral parameters of the laboratory spectra of meteorites (Table 3 and Table 4) in the Band I center vs. BAR plot in Fig. 12. Most of the HED meteorites fall in the basaltic achondrites region; the ordinary chondrites are clearly divided in three zones within the S(IV) polygon, and the olivine-dominated meteorites such as R chondrite, brachinite, and CV chondrite fall above the olivine-orthopyroxene mixing line. Most of the ureilites are found clustered near the S(III) zone, whereas the lodranites and acapulcoites are more scattered. The olivine-rich unit plots in the S(IV) region, slightly below the olivine-orthopyroxene mixing line as defined in Gaffey et al. (1993). Asteroids falling in the S(IV) region could be analogous to ordinary chondrite meteorites. Dunn et al. (2010b) further divided the S(IV) region to identify the locations of the H, L and LL chondrite subtypes. The olivine-rich unit observed by VIR falls in the L chondrite group within the S(IV) region of the plot near the cluster corresponding to ureilite meteorites near the S(III) zone. This suggests that the origin of the olivine on Vesta could be from the contamination of the regolith by ordinary chondrites (possibly L), ureilites or mixtures of HED with some olivine-dominated meteorites such as R-chondrites.

*4.3 Meteoritical Evidence from HEDs*

Olivine-bearing HEDs are rare and are mostly found among diogenites, even though some howardites can contain a small amount olivine (< few %) (Beck et al., 2012). In the Meteoritical Bulletin Database, 11 diogenites (some of them are paired) among 379 diogenites are classified as "diogenite-olivine" (i.e. olivine-rich diogenites). The olivine modal abundance ranges from 10 vol.% to 50 vol.% for these olivine-rich diogenites. Thangjam et al. (2014) did a literature survey and found about 30 diogenites have < 10 vol.% olivine, 8 diogenites have 40 to 68 vol.% olivine and 4 dunites have > 90 vol.% olivine (Bunch et al., 2006; Bunch et al., 2010; Beck et al., 2011). Desnoyers and Jerome (1973) reported rare isolated grains in 8 howardites, usually heterogeneously distributed, and suggested an exogenous origin. Delaney et al. (1980) found an average modal content of 0.5% olivine in 9 howardites. According to Bischoff et al., (2006) regolith mixing was more limited on Vesta than for carbonaceous asteroids because of the lack of olivine-rich mantle material in howardites. This assumed that abundant olivine was excavated in the first place during the formation of Rheasilvia basin, which does not seem to be the case. In addition, Sack et al. (1991) concluded from the limited number of olivine-rich diogenites that only the upper layers of the HED parent body have been excavated.

Analysis of Dawn mission data has demonstrated that impact gardening is prevalent on Vesta (Pieters et al., 2012) and that there is evidence for contamination by carbonaceous chondrites in the vestan regolith (e.g., Reddy et al., 2012a, Prettyman et al., 2012). Therefore, contamination by another meteorite type, for example ordinary chondrite, should not come as a surprise although it is not as widespread (or as extensively documented) as carbonaceous chondrite material in howardites. A good example of exogenous material in a howardite, in this case ordinary chondrite, has been recently published by Janots et al. (2012). According to this work, chondritic clasts were incorporated in the impact melt of JaH 556 (before or during a major impact event) with partial dissolution of the non-Vesta material in a matrix composed of eucritic and diogenitic components. Similar clast incorporation and clast-melt interaction phenomena are widely observed in melt-bearing impactite lithologies from terrestrial impact structures (e.g., Grieve, 1987). Therefore, JaH 556 could have formed by incorporation of ordinary chondrite impactor remnants in the ejecta blanket of an impact crater and subsequent excavation of this regolith material from Vesta by another impact with enough velocity to allow ejection into space. A similar scenario could be considered for the exogenous olivine observed by VIR at Arruntia and Bellicia craters, with contamination of crater walls and ejecta blankets with remnants of impactor material rich in olivine. Although JaH 556 is thus far unique in the HED collection, one might expect a significant contribution of ordinary chondrite material (millimetric or larger size clasts) on Vesta's surface assuming the meteorite flux is similar to the Earth's (Janots et al., 2012).

It is interesting to note that ordinary chondrites have also been suggested to play a role in the formation of Vesta as one of the precursor material. Boesenberg and Delaney (1997) determined that the probable bulk composition of Vesta could be modeled as a mixture of ordinary chondrites (70 wt.% H chondrites) and carbonaceous chondrites (30 wt.% CM chondrites) based on oxygen isotope constrains, Fe/Mn ratio and partial melting experiments. Righter and Drake (1997) proposed a model using a mixture of CM and L chondrites based on similar criteria. More recently, Toplis et al. (2013) proposed a Na-depleted H-chondrite precursor mixed with ~ 25 % of CM-like material as a possible Vesta analogue. If true, this would imply that ordinary chondrite materials are found in the vicinity of Vesta and the same material that originally accreted to form Vesta could contribute to late impactors. Each of the Dawn scientific instruments found evidence for extensive contamination of the vestan surface by carbonaceous material likely brought by micrometeorites bombardment and/or by the low-velocity impact event at the Veneneia impact basin (De Sanctis et al., 2013; Prettyman et al., 2012; Reddy et al., 2012a; Turrini et al., 2014). Alternatively, Consolmagno et al. (2014) proposed different scenarios to account for the density and surface composition findings of Dawn observations of Vesta: 1) either Vesta's precursor material corresponds to a non-chondritic bulk material, or 2) an object three times the mass of Vesta underwent a violent collision during which the object was disrupted, and its primordial olivine and iron were removed as it reaccreated to form proto-Vesta. In both options, olivine from Vesta would not be exposed at the surface. Therefore, these models would explain the lack of widespread olivine-rich terrains identified on Vesta, and, consequently, any detection of local olivine-enriched material would have to be extraneous contributions from impactors.

## 5. Summary

Our extensive analysis of spectral data from the VIR instrument of olivine-rich units in the northern hemisphere of Vesta has provided valuable insights into the nature and origin of this material. While the existence of several olivine-rich HED meteorites clearly demonstrates that Vesta contains endogenic olivine-rich rocks, the endogenic hypothesis for the olivine-rich unit is not fully compatible with the observations of the Dawn mission. In particular, if the observed olivine was excavated during the formation of the 19 km deep Rheasilvia basin, then it should be ubiquitous both on the surface and in HED meteorites. Analysis of Dawn spectrometer data, however, shows that olivine-rich units are not widespread in the Rheasilvia basin (Ammannito et al., 2013a and 2013b; Clenet et al., 2014), and olivine-rich HEDs and dunites are rare. Clenet et al. (2014) argued that the vestan crust is thicker than predicted by most current models, and therefore that the Moho (crust-mantle boundary) was not breached during the formation of the Rheasilvia basin. If the olivine detected by Ammannito et al. (2013a) in the northern hemisphere of Vesta was formed in plutons and the observed olivine-rich units are impact excavated plutons, then similar occurrences of exposed plutonic olivine-rich material should be found everywhere on Vesta. Interestingly, Wilson and Keil (1996) suggested that large-scale intrusions should not be common on Vesta. Excavation of dikes in the size range suggested by Wilson and Keil (1996) would likely produce much more limited exposures of olivine that would be difficult to detect with the VIR instrument (Beck et al., 2013). The fact that the olivine would likely be at the bottom of these intrusive bodies also limit the detectability.

The association of olivine with small to medium impact craters in the region encompassing Bellicia and Arruntia suggests an exogenic origin rather than exposed mantle or magmatic intrusions. If this is true, here is what we know:

- Ground-based and HST observations of Vesta have suggested the presence of olivine (Binzel et al., 1997; Gaffey et al., 1997). These observations were reinforced by the discovery of olivine-rich HED meteorites from Vesta in recent years (Beck et al., 2013). However, analysis of Dawn data has shown that this "olivine-bearing unit" is most likely impact melt in the ejecta of Oppia crater (Le Corre et al., 2013; Reddy et al., 2013).
- The lack of widespread mantle olivine, exposed during the formation of the 19 km deep (Schenk et al., 2012) Rheasilvia basin on Vesta's South Pole, further complicates this picture (Beck et al. 2013; Clenet et al. 2014).
- Ammannito et al., (2013a) reported the discovery of local-scale olivine-rich units concentrated in the walls and ejecta of Arruntia (10.5 km in diameter) and Bellicia (41.7 km in diameter) craters located in the northern hemisphere, 350-430 km from the Rheasilvia basin. They suggest exposed plutons as a probable source of these olivine-rich units but still favor the magma ocean model because of the large size of the nearly pure olivine exposures.
- Our mineralogical analysis using diagnostic spectral band parameters gives olivine abundance between 70-80 vol.%, which is consistent with those obtained by Ammannito et al. (2013a) using non-diagnostic curve match techniques. They estimated an olivine concentration between 50-80 vol.%, however, no single olivine + pyroxene mixture matches the spectrum of the olivine-rich unit.
- Meteoritic evidence suggests contamination of HEDs by various exogenic materials rich in olivine such as primitive achondrites (ureilites, lodranites, acapulcoites), pallasites and ordinary chondrites (including H, L and LL chondrites). This includes the howardite JaH 556, which contains ~20 vol.% H chondrite material mixed with HED impact melt.

- Combining non-diagnostic curve matching with mineralogical analysis using diagnostic spectral band parameters, we suggest that the olivine-rich material in the northern hemisphere of Vesta could also be explained by the delivery of exogenic material. This exogenic material could include pure ordinary chondrites or mixtures of HED with either ordinary chondrites or olivine-dominated meteorites such as R-chondrites.
- If the observed olivine on the vestan surface is largely attributable to exogenic sources, this favors the model of Clenet et al. (2014) with a very deep crust-mantle boundary on Vesta.

Even if H chondrite xenoliths have been found in howardites with abundances of up to ~20 vol.% in JaH 556 (Janots et al., 2012), the amount of meteoritic fragments made of ordinary chondrites and differentiated meteorites correspond to ~1% of all xenoliths typically found in HED breccias (Lorenz et al., 2007). Given the limited data set analyzed in this study, a more thorough investigation looking at multiple olivine-rich sites on Vesta and using more robust laboratory spectral library is warranted. Our work here attempts to lay out the broad outline of the hypothesis and further research is strongly encouraged.

**7. Acknowledgements**

We would like to thank Paul Buchanan and an anonymous reviewer for their useful comments and discussion in improving the manuscript. This research utilizes spectra acquired by Richard P. Binzel, Takahiro Hiroi, Tim McCoy, Harry Y. McSween, David W. Mittlefehldt, Carle M. Pieters, and Allan H. Treiman with the NASA RELAB facility at Brown University. This research work was supported by NASA Planetary Mission Data Analysis Program Grant NNX14AN16G, NASA NEOO Program Grant NNX14AL06G, and NASA Planetary Geology and Geophysics Grant NNX14AN35G. EAC wishes to thank CFI, MRIF, CSA, and the University of Winnipeg for their support in establishing the University of Winnipeg Planetary Spectrophotometer Facility, and CSA and NSERC for supporting this study. Finally, we thank Michael Farmer for helping procure some of the meteorite samples for this study.

## Tables

**Table 1.** Band parameters and associated errors for the olivine-orthopyroxene mixtures.

| Mixtures | Band I Center (µm) | BI Center 1-σ | Band II Center (µm) | BII Center 1-σ | BAR | BAR 1-σ |
|---|---|---|---|---|---|---|
| 100 % opx | 0.9134 | 0.0014 | 1.8581 | 0.0027 | 1.8210 | 0.1711 |
| 10 % olv + 90 % opx | 0.9131 | 0.0016 | 1.8651 | 0.0072 | 1.5947 | 0.2319 |
| 20 % olv + 80 % opx | 0.9133 | 0.0004 | 1.8694 | 0.0014 | 1.3910 | 0.1974 |
| 30 % olv + 70 % opx | 0.9149 | 0.0015 | 1.8738 | 0.0023 | 1.1144 | 0.1352 |
| 40 % olv + 60 % opx | 0.9165 | 0.0002 | 1.8811 | 0.0022 | 0.9311 | 0.2400 |
| 50 % olv + 50 % opx | 0.9173 | 0.0043 | 1.8868 | 0.0034 | 0.9311 | 0.0958 |
| 60 % olv + 40 % opx | 0.9222 | 0.0074 | 1.8958 | 0.0036 | 0.7430 | 0.0457 |
| 70 % olv + 30 % opx | 0.9317 | 0.0226 | 1.9064 | 0.0054 | 0.5241 | 0.0783 |
| 80 % olv + 20 % opx | 0.9831 | 0.0171 | 1.923 | 0.0077 | 0.3240 | 0.0933 |
| 90 % olv + 10 % opx | 1.0424 | 0.0032 | 1.9274 | 0.0059 | 0.1662 | 0.0354 |
| 100 % olv | 1.0555 | 0.0013 | 1.941 | 0.0026 | 0.0173 | 0.0061 |

**Table 2.** Olivine abundance (wt.%) for different types of olivine-bearing meteorites and summary of the presence of exogenous material from the literature (e.g. Lorenz et al., 2007; Janots et al., 2012; Beck et al., 2012). Table adapted from Cloutis et al. (in press). Olivine content is from Dunn et al. (2010a) for the ordinary chondrites, from Howard et al. (2009) for the CV chondrites, from Mittlefehldt et al. (2003) for brachinites. Other values are from Cloutis et al. (in press).

| Meteorite class | Olivine abundance | Evidence in HEDs |
|---|---|---|
| *Primitive meteorites* | | |
| **CV chondrites** | 75-85 | Maybe |
| **R chondrites** | 70-90 | No |
| *Differentiated meteorites* | | |
| **Angrites** | 0-64 | No |
| **Brachinites** | 80-90 | No |
| **Acapulcoites/Lodranites** | Variable | Maybe |
| **Pallasites** | 25-75 | Maybe |
| **Ureilites** | Variable | Yes |
| **Ordinary chondrites** | H: 29-40<br>L: 38-45<br>LL: 47-57 | Yes |

**Table 3.** List of meteorites from the RELAB database with their band parameters used for the comparison with the VIR data.

| Sample | RELAB ID | RELAB file name | Type | Band I Center (microns) | Band II Center (microns) | BAR |
|---|---|---|---|---|---|---|
| Bernares (a) | MT-HYM-083 | C1MT83 | LL4 | 1.008 | 1.984 | 0.522 |
| Greenwell Springs | TB-TJM-075 | C1TB75 | LL4 | 0.991 | 1.959 | 0.512 |
| Hamlet | MT-HYM-075 | C1MT75 | LL4 | 0.980 | 1.974 | 0.561 |
| Witsand Farm | MT-HYM-076 | C1MT76 | LL4 | 1.004 | 2.001 | 0.457 |
| Aldsworth | MT-HYM-077 | C1MT77 | LL5 | 0.977 | 1.960 | 0.561 |
| Alta'ameem | MT-HYM-078 | C1MT78 | LL5 | 0.986 | 1.952 | 0.435 |
| Olivenza | MT-HYM-085 | C1MT85 | LL5 | 1.012 | 1.969 | 0.387 |
| Paragould | MT-HYM-079 | C1MT79 | LL5 | 0.985 | 1.968 | 0.409 |
| Tuxtuac | MT-HYM-080 | C1MT80 | LL5 | 1.035 | 1.950 | 0.310 |
| Bandong | TB-TJM-067 | C1TB67 | LL6 | 1.001 | 1.987 | 0.274 |
| Cherokee Springs | TB-TJM-090 | C1TB90 | LL6 | 0.989 | 1.940 | 0.406 |
| Karatu | TB-TJM-077 | C1TB77 | LL6 | 1.006 | 1.992 | 0.311 |
| Saint-Severin | TB-TJM-145 | C1TB145 | LL6 | 1.003 | 1.881 | 0.269 |
| Attarra | TB-TJM-065 | C1TB65 | L4 | 0.957 | 1.927 | 0.726 |
| Bald Mountain | TB-TJM-102 | C1TB102 | L4 | 0.929 | 2.001 | 0.960 |
| Rio Negro | TB-TJM-081 | C1TB81 | L4 | 0.953 | 1.932 | 0.813 |
| Rupota | TB-TJM-121 | C1TB121 | L4 | 0.955 | 1.951 | 0.582 |
| Ausson | MT-HYM-084 | C1MT84 | L5 | 0.930 | 1.918 | 1.031 |
| Blackwell | MT-HYM-081 | C1MT81 | L5 | 0.955 | 1.953 | 0.526 |
| Cilimus | MT-HYM-082 | C1MT82 | L5 | 0.950 | 1.925 | 0.515 |
| Guibga | TB-TJM-134 | C1TB134 | L5 | 0.962 | 1.943 | 0.633 |
| Mabwe-Khoywa | TB-TJM-107 | C1TB107 | L5 | 0.954 | 1.944 | 0.744 |
| Malakal | TB-TJM-109 | C1TB10 | L5 | 0.954 | 1.977 | 0.709 |
| | | 9 | | | | |
| Messina | TB-TJM-099 | C1TB99 | L5 | 0.959 | 1.960 | 0.595 |
| Apt | TB-TJM-064 | C1TB64 | L6 | 0.963 | 1.940 | 0.511 |
| Aumale | TB-TJM-101 | C1TB101 | L6 | 0.960 | 1.946 | 0.614 |
| Karkh | TB-TJM-137 | C1TB137 | L6 | 0.964 | 1.930 | 0.371 |
| Kunashak | TB-TJM-139 | C1TB139 | L6 | 0.972 | 1.942 | 0.518 |
| Kyushu | TB-TJM-140 | C1TB140 | L6 | 0.970 | 1.933 | 0.544 |
| New Concord | TB-TJM-130 | C1TB130 | L6 | 0.960 | 1.928 | 0.606 |
| Farmville | TB-TJM-128 | C1TB128 | H4 | 0.939 | 1.928 | 0.789 |
| Forest Vale | TB-TJM-093 | C1TB93 | H4 | 0.937 | 1.934 | 0.800 |
| Kabo | TB-TJM-136 | C1TB136 | H4 | 0.939 | 1.934 | 0.915 |
| Marilia | TB-TJM-078 | C1TB78 | H4 | 0.934 | 1.945 | 0.851 |
| Sao Jose do Rio Preto | TB-TJM-082 | C1TB82 | H4 | 0.942 | 1.901 | 0.992 |
| Allegan | TB-TJM-104 | C1TB104 | H5 | 0.930 | 1.897 | 1.068 |
| Ehole | TB-TJM-074 | C1TB74 | H5 | 0.940 | 1.945 | 0.999 |
| Itapicuru-Mirim | TB-TJM-097 | C1TB97 | H5 | 0.944 | 1.922 | 0.878 |
| Lost City | TB-TJM-129 | C1TB129 | H5 | 0.940 | 1.936 | 0.857 |
| Primbram | TB-TJM-143 | C1TB143 | H5 | 0.940 | 1.906 | 0.857 |
| Schenectady | TB-TJM-083 | C1TB83 | H5 | 0.937 | 1.913 | 0.786 |
| Uberaba | TB-TJM-085 | C1TB85 | H5 | 0.945 | 1.943 | 0.881 |
| Andura | TB-TJM-088 | C1TB88 | H6 | 0.927 | 1.929 | 0.889 |
| Bustura | TB-TJM-069 | C1TB69 | H6 | 0.936 | 1.918 | 0.795 |
| Canon City | TB-TJM-131 | C1TB131 | H6 | 0.953 | 1.932 | 0.752 |
| Chiang Khan | TB-TJM-132 | C1TB132 | H6 | 0.946 | 1.920 | 0.846 |
| Guarena | TB-TJM-094 | C1TB94 | H6 | 0.938 | 1.916 | 0.748 |
| Ipiranga | TB-TJM-135 | C1TB135 | H6 | 0.939 | 1.921 | 0.835 |
| PCA 82506 <500 microns | MC-RPB-006 | c1mc06 | Ureilite | 0.930 | 1.928 | 1.014 |
| NWA 1500 <75 microns | PH-D2M-018 | c1ph18 | Ureilite | 1.076 | 2.120 | 0.177 |
| Y-791538 <75 microns | PH-D2M-019 | c1ph19 | Ureilite | 0.928 | 1.989 | 0.273 |
| META 78008 <75 microns | PH-D2M-020 | c1ph20 | Ureilite | 1.027 | | |
| ALH 82130 <75 microns | PH-D2M-021 | c1ph21 | Ureilite | 0.938 | | |
| LEW 88201 <75 microns | PH-D2M-022 | c1ph22 | Ureilite | 0.950 | | |
| GRA 95205 <75 microns | PH-D2M-023 | c1ph23 | Ureilite | 0.930 | 1.974 | 0.482 |
| EET 87517 <75 microns | PH-D2M-024 | c1ph24 | Ureilite | 0.921 | 1.930 | 0.429 |
| GRO 95575 <75 microns | PH-D2M-025 | c1ph25 | Ureilite | 0.950 | 2.036 | 0.284 |
| PCA 82506 <75 microns | PH-D2M-026 | c1ph26 | Ureilite | 0.948 | 1.934 | 0.320 |
| ALHA 77257 <75 microns | PH-D2M-027 | c1ph27 | Ureilite | 0.931 | 1.905 | 0.309 |
| Novo-Urei <75 microns | PH-D2M-028 | c1ph28 | Ureilite | 0.951 | | |
| Goalpara <75 microns | PH-D2M-029 | c1ph29 | Ureilite | 0.946 | | |
| ALHA 81101 <75 microns | PH-D2M-030 | c1ph30 | Ureilite | 0.952 | 1.990 | 0.218 |
| EET 96042 <75 microns | PH-D2M-031 | c1ph31 | Ureilite | 0.959 | 1.993 | 0.196 |
| Y-791538 <75 microns2 | PH-D2M-036 | c1ph36 | Ureilite | 0.927 | 1.974 | 0.218 |
| EET 87720 <75 microns | PH-D2M-037 | c1ph37 | Ureilite | 0.925 | 1.940 | 0.337 |
| Kenna <75 microns | PH-D2M-038 | c1ph38 | Ureilite | 0.972 | 2.024 | 0.273 |
| DaG 319 <75 microns | PH-D2M-039 | c1ph39 | Ureilite | 0.957 | 1.979 | 0.228 |
| MET 01085 <75 microns | PH-D2M-040 | c1ph40 | Ureilite | 0.923 | 1.880 | 0.665 |
| Y-791538 <125 microns | MP-TXH-101 | c2mp101 | Ureilite | 0.931 | 1.932 | 0.309 |
| Y-74659 <25 microns | MB-TXH-087-A | camb87 | Ureilite | 0.926 | | |
| Y-74659 25-45 microns | MB-TXH-087-B | cbmb87 | Ureilite | 0.920 | 1.879 | 0.750 |
| Y-74659 45-75 microns | MB-TXH-087-C | ccmb87 | Ureilite | 0.934 | 1.906 | 0.387 |
| Lodran <1000 microns | MB-CMP-017 | c1mb17 | Lodranite | 0.963 | 1.908 | 0.987 |
| ALHA 81261 <125 microns | TB-TJM-039 | c1tb39 | Acapulcoite | 0.934 | 1.875 | 0.702 |
| ALHA 81187 <125 microns | TB-TJM-040 | c1tb40 | Acapulcoite | 0.930 | 1.943 | 0.325 |
| Lodran <125 microns | TB-TJM-041 | c1tb41 | Lodranite | 0.928 | 1.888 | 0.937 |
| EET 84302 | TB-TJM-042 | c1tb42 | Acapulcoite/Lodranite | 0.922 | 1.898 | 1.359 |
| Acapulco | TB-TJM-043 | c1tb43 | Acapulcoite | 0.931 | 1.885 | 0.872 |
| GRA 95209 | TB-TJM-044 | c1tb44 | Lodranite | 0.924 | 1.915 | 0.831 |
| LAP 04840 <125 microns | DD-AHT-107 | c1dd107 | R chondrite | 1.030 | 1.963 | 0.030 |
| LAP 04840 <45 microns | DD-AHT-108 | c1dd108 | R chondrite | 1.046 | 1.960 | 0.025 |
| PCA 91002 <63 microns | MB-TXH-065-A | Camb65 | R chondrite | 1.068 | 2.184 | 0.227 |
| PCA 91002 63-125 microns | MB-TXH-065-B | Cbmb65 | R chondrite | 1.062 | 2.104 | 0.216 |
| PCA 91002 <125 microns | MB-TXH-065 | c1mb65 | R chondrite | 1.060 | 2.047 | 0.155 |
| A-881988 <125 microns | MP-TXH-059 | c1mp59 | R chondrite | 1.052 | 1.964 | 0.033 |
| Rumuruti <150 microns | MT-TJM-013 | c1mt13 | R chondrite | 1.068 | 2.120 | 0.061 |
| EET 96026 <125 microns | MT-TXH-014 | c1mt14 | R chondrite | 1.072 | 2.076 | 0.495 |
| EET 96026 <125 microns | MT-TXH-014 | c2mt14 | R chondrite | 1.048 | 2.151 | 0.440 |
| PRE 95411 | MT-TXH-045 | c1mt45 | R chondrite | 1.067 | 1.976 | 0.293 |
| NWA 753 <125 microns | TB-TJM-114 | c1tb114 | R chondrite | 1.060 | 1.956 | 0.077 |
| EET 99402 <125 microns | TB-TJM-058 | c1tb58 | Brachinite | 1.072 | 2.090 | 0.015 |
| QUE 93744 <75 microns | PH-D2M-041 | c1ph41 | CV chondrite | 1.064 | 2.017 | 0.830 |

**Table 4.** List of HED meteorites with their band parameters used for the comparison with the VIR data.

| Meteorite | Group | Subtype | Band I Center (microns) | Band II Center (microns) | BAR |
|---|---|---|---|---|---|
| A-881526 | diogenite | polymict | 0.920 | 1.896 | 1.865 |
| EETA79002 | diogenite | | 0.920 | 1.891 | 1.654 |
| GRO 95555 | diogenite | anomalous | 0.923 | 1.912 | 1.963 |
| Johnstown | diogenite | | 0.918 | 1.887 | 1.523 |
| LAP 91900 | diogenite | | 0.923 | 1.909 | 1.907 |
| Y-74013 | diogenite | | 0.924 | 1.920 | 1.859 |
| Y-75032 | diogenite | Type B | 0.927 | 1.941 | 1.636 |
| ALHA77256 | diogenite | | 0.922 | 1.894 | 2.157 |
| ALH-78132 | eucrite | polymict | 0.928 | 1.950 | 1.872 |
| Serra de Magé | eucrite | cumulate | 0.930 | 1.967 | 1.741 |
| Moore County | eucrite | cumulate | 0.938 | 1.993 | 1.621 |
| A-881819 | eucrite | cumulate | 0.932 | 1.966 | 1.808 |
| EETA79005 | eucrite | polymict | 0.934 | 1.959 | 1.999 |
| EETA79006 | eucrite | polymict | 0.936 | 1.970 | 2.176 |
| LEW 87004 | eucrite | polymict | 0.936 | 1.978 | 1.678 |
| Pasamonte | eucrite | polymict | 0.940 | 2.014 | 1.728 |
| PCA 91006 | eucrite | brecciated | 0.942 | 1.993 | 1.675 |
| ALH 85001 | eucrite | cumulate | 0.933 | 1.957 | 1.879 |
| Juvinas | eucrite | monomict | 0.937 | 2.001 | 1.705 |
| PCA 82501 | eucrite | unbrecciated | 0.943 | 1.998 | 2.272 |
| PCA 91007 | eucrite | brecciated | 0.943 | 2.029 | 2.439 |
| Petersburg | eucrite | polymict | 0.935 | 1.975 | 2.135 |
| Y-74450 | eucrite | polymict | 0.936 | 1.975 | 1.398 |
| ALHA81001 | eucrite | monomict | 0.937 | 1.966 | 1.816 |
| PCA 82502 | eucrite | unbrecciated | 0.942 | 2.029 | 2.377 |
| LEW 85303 | eucrite | polymict | 0.945 | 2.030 | 1.715 |
| EET 92003 | eucrite | brecciated | 0.933 | 2.004 | 1.921 |
| Millbillillie | eucrite | monomict | 0.939 | 2.013 | 1.593 |
| Stannern | eucrite | monomict | 0.938 | 2.015 | 2.118 |
| Jonzac | eucrite | monomict | 0.940 | 2.010 | 1.992 |
| Béréba | eucrite | monomict | 0.942 | 2.025 | 1.778 |
| Cachari | eucrite | monomict | 0.941 | 2.003 | 1.750 |
| GRO 95533 | eucrite | brecciated | 0.940 | 2.023 | 1.755 |
| EET 87542 | eucrite | brecciated | 0.940 | 2.006 | 1.553 |
| EET 83376 | howardite | | 0.935 | 1.958 | 1.940 |
| EET 87503 | howardite | | 0.931 | 1.954 | 1.499 |
| EET 87513 | howardite | | 0.934 | 1.967 | 1.557 |
| GRO 95535 | howardite | | 0.931 | 1.962 | 1.906 |
| GRO 95574 | howardite | | 0.929 | 1.953 | 1.967 |
| Kapoeta | howardite | | 0.930 | 1.987 | 1.324 |
| Le Teilleul | howardite | | 0.929 | 1.945 | 2.025 |
| QUE 94200 | howardite | | 0.924 | 1.928 | 2.031 |
| QUE 97001 | howardite | | 0.924 | 1.931 | 2.313 |
| Y-7308 | howardite | | 0.928 | 1.946 | 1.976 |
| Y-790727 | howardite | | 0.931 | 1.959 | 2.069 |
| Y-791573 | howardite | | 0.928 | 1.945 | 2.022 |

## Captions

**Figure 1.** Spectral data from the Dawn/VIR instrument of the olivine-rich site from the Bellicia crater plotted with olivine-orthopyroxene mixtures. The plot has been modified from Ammannito et al. (2013a). VIR data were digitized from Ammannito et al. (2013a) and olivine-orthopyroxene mixtures are from HOSERlab (continuum is removed with a polynomial). The grain size range 0-38 μm is selected because it appears to reproduce the shape of the VIR spectrum better than the other grain size ranges existing in the HOSERlab database. This is reasonable, as fine-grained material is expected to dominate the scattering from the vestan regolith (e.g., Hiroi et al., 1995).

**Figure 2.** Similar plot to figure 1 but with selected mixtures of olivine and orthopyroxene, which correspond to the range of olivine content suggested in Ammannito et al. (2013a) for the olivine-rich unit in Bellicia crater.

**Figure 3.** Continuum-removed spectra of **(a)** all the RELAB spectra of acapulcoites and lodranites, and **(b)** the one we selected as best fit to the VIR data of the olivine-rich unit.

**Figure 4.** Continuum-removed spectra of **(a)** all the RELAB spectra of ureilites, and **(b)** the one we selected as best fit to the VIR data of the olivine-rich unit.

**Figure 5.** Continuum-removed spectra of (**a)** all the RELAB spectra of H chondrites, **(b)** all the RELAB spectra of L chondrites, **(c)** all the RELAB spectra of LL chondrites, **(d)** selected H chondrites that best fit the VIR data, **(e)** selected L chondrites that best fit the VIR data, **(f)** selected LL chondrites that best fit the VIR data.

**Figure 6.** Continuum-removed spectra of R-chondrites from RELAB.

**Figure 7.** Continuum-removed spectra of pallasites from RELAB.

**Figure 8.** Example of continuum-removed spectrum of CV chondrites from RELAB.

**Figure 9.** Continuum-removed spectrum of brachinite from RELAB.

**Figure 10.** Continuum-removed spectra of **(a)** selected mixtures of H, L and LL chondrites with the eucrite NWA 7465, **(b)** selected mixtures of R-chondrite Rumuruti, pallasites Thiel Mountain and brachnite EET99402 with the eucrite NWA 7465.

**Figure 11. a)** Plot of Band I center vs. Band II center showing all meteorites we obtained from RELAB and including HED meteorites. **b)** Plot of Band I center vs. Band II center with zones for the H, L, and LL chondrites and an ellipse defining the area for the HEDs.

Black circle with error bars corresponds to the VIR data from the olivine-rich unit at the Bellicia crater as observed by Ammannito et al. (2013a). For the error bar of the BI and BII centers of the olivine-rich unit seen by VIR we used possible variations of the band centers within the temperature range observed on Vesta. The effect of changing temperature on the spectra of minerals, meteorites and asteroids has been well documented (e.g., Singer and Roush, 1985; Moroz et al., 2000; Hinrichs and Lucey; 2002; Reddy et al., 2012b; Sanchez et al., 2014). With an increase or decrease in temperature, band centers shift to longer or shorter wavelength and absorption bands expand or contract. Temperature ranges for dark material and bright materials units on Vesta are presented in the work of Tosi et al. (2014). Average temperatures are, for the dark material, 252-269 K, and 245-255 K for the bright material. The minimum temperature is ~200 K during dawn or dusk and the maximum is ~270 K. Therefore, we decided to compute temperature corrections for the range 200-280 K. Because pyroxene is likely a dominant mineral in the assemblage we used HED temperature correction equations from Reddy et al. (2012b). Using HED equations, the error bar for BI center is +/-0.0022 and for BII center +/-0.0068. BI and BII centers for VIR data are not corrected for temperature.

**Figure 12.** Plot of Band I center vs. BAR for the meteorites we obtained from RELAB. The olivine-rich site from Bellicia crater is represented by a black filled circle. The plot is adapted from Gaffey et al. (1993) and shows the S-asteroid subtypes. Most HEDs plot in the basaltic achondrites (BA) rectangle and ordinary chondrites plot in the boot-shaped S(IV) region. Pure olivine plots in the S(I) box to the top left where the BAR is close to zero. The black curve corresponds to the olivine-orthopyroxene mixing line from Cloutis et al. (1986). Dashed lines are delimiting the H, L and LL chondrite zones (Dunn et al., 2010b). The error bar of the BAR parameter for the olivine-rich unit observed by VIR is assessed by deriving the BAR values ten times using our Matlab code and computing 2-sigma error.

**Figures**

Fig. 1

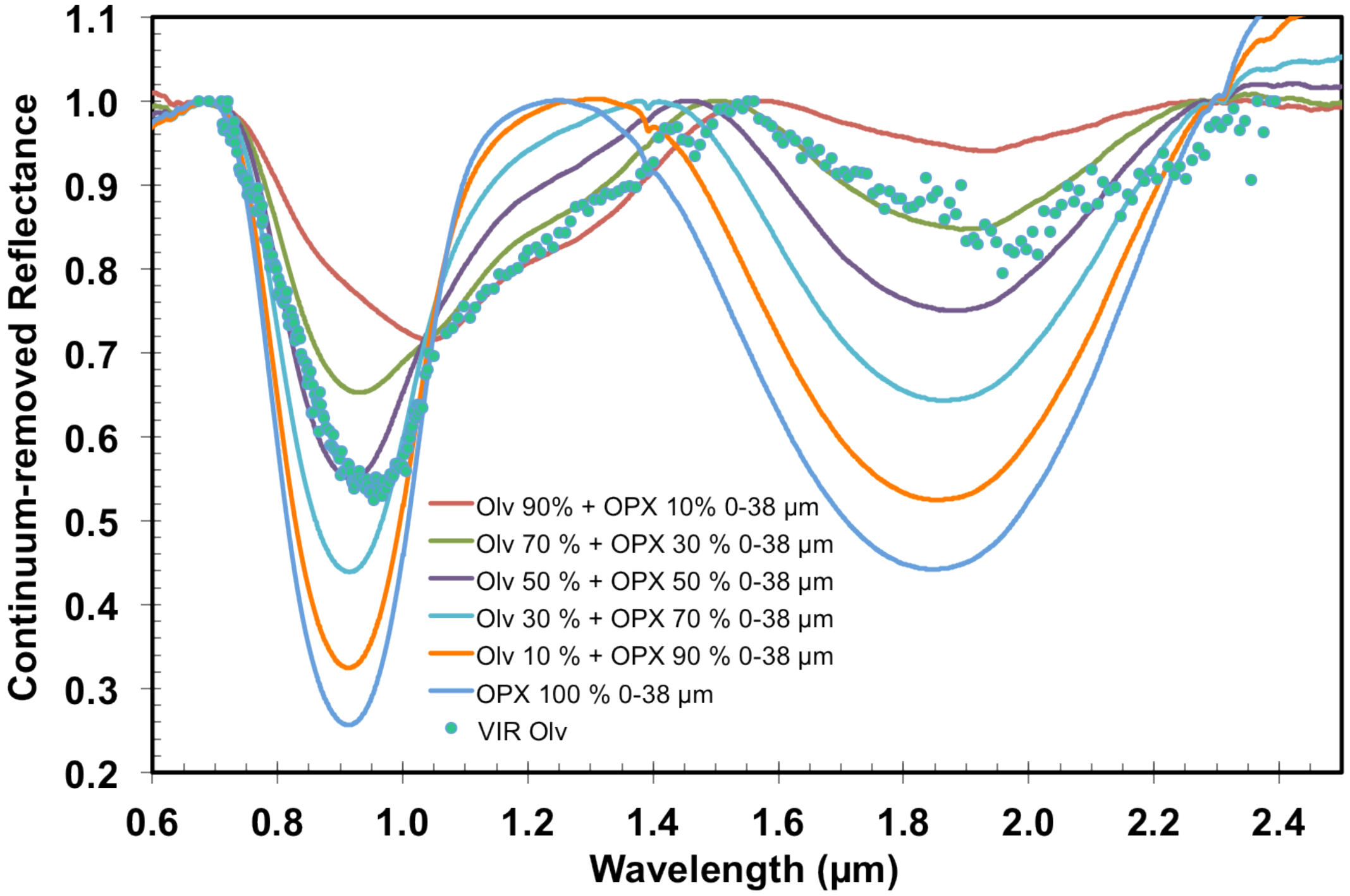

Fig. 2

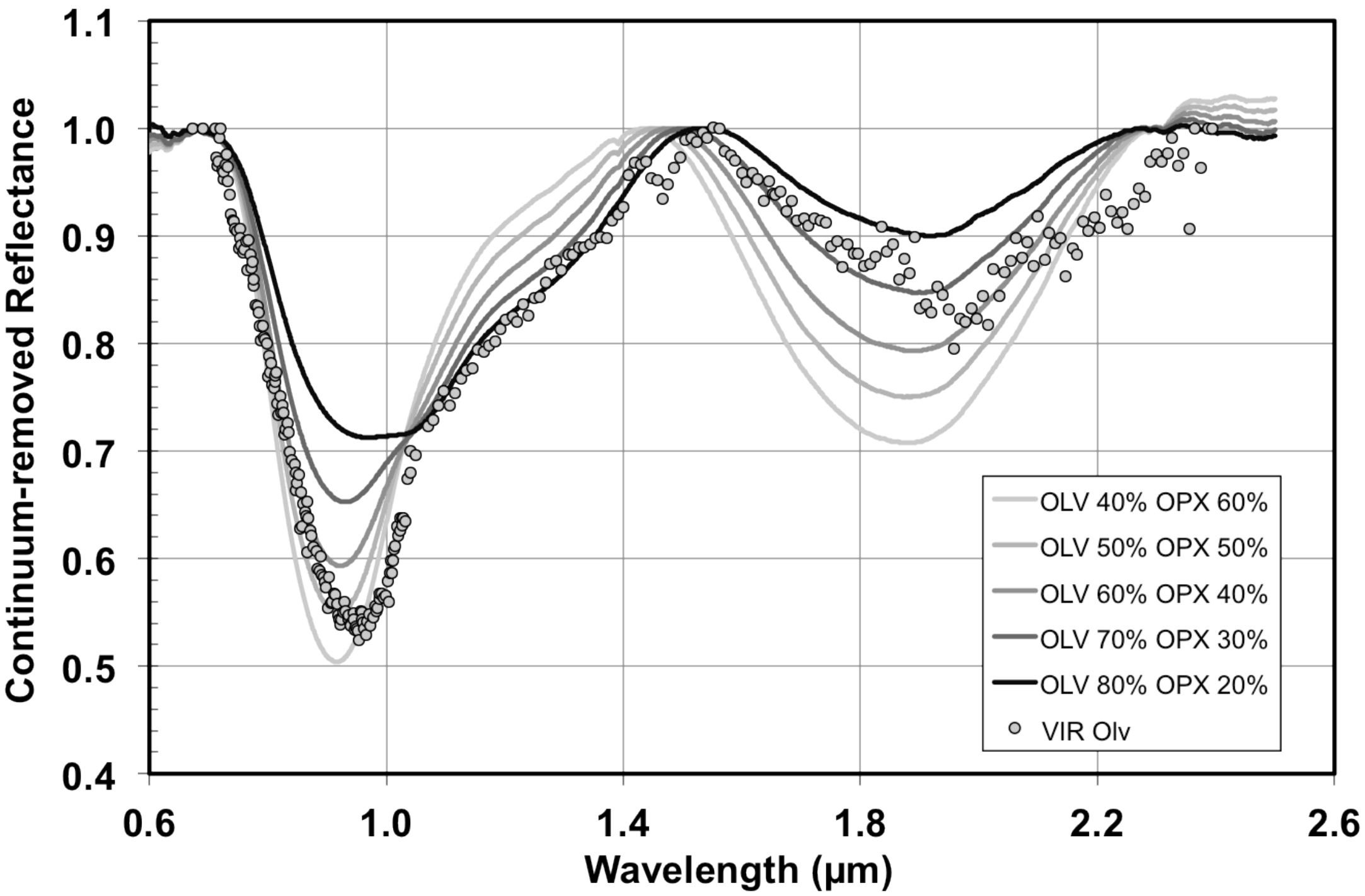

Continuum-removed Reflectance
1.1
1.0
0.9
0.8
0.7
0.6
0.5
0.4
0.6
1.0
1.4
1.8
2.2
2.6
Wavelength (µm)
OLV 40% OPX 60%
OLV 50% OPX 50%
OLV 60% OPX 40%
OLV 70% OPX 30%
OLV 80% OPX 20%
VIR Olv

Fig. 3a

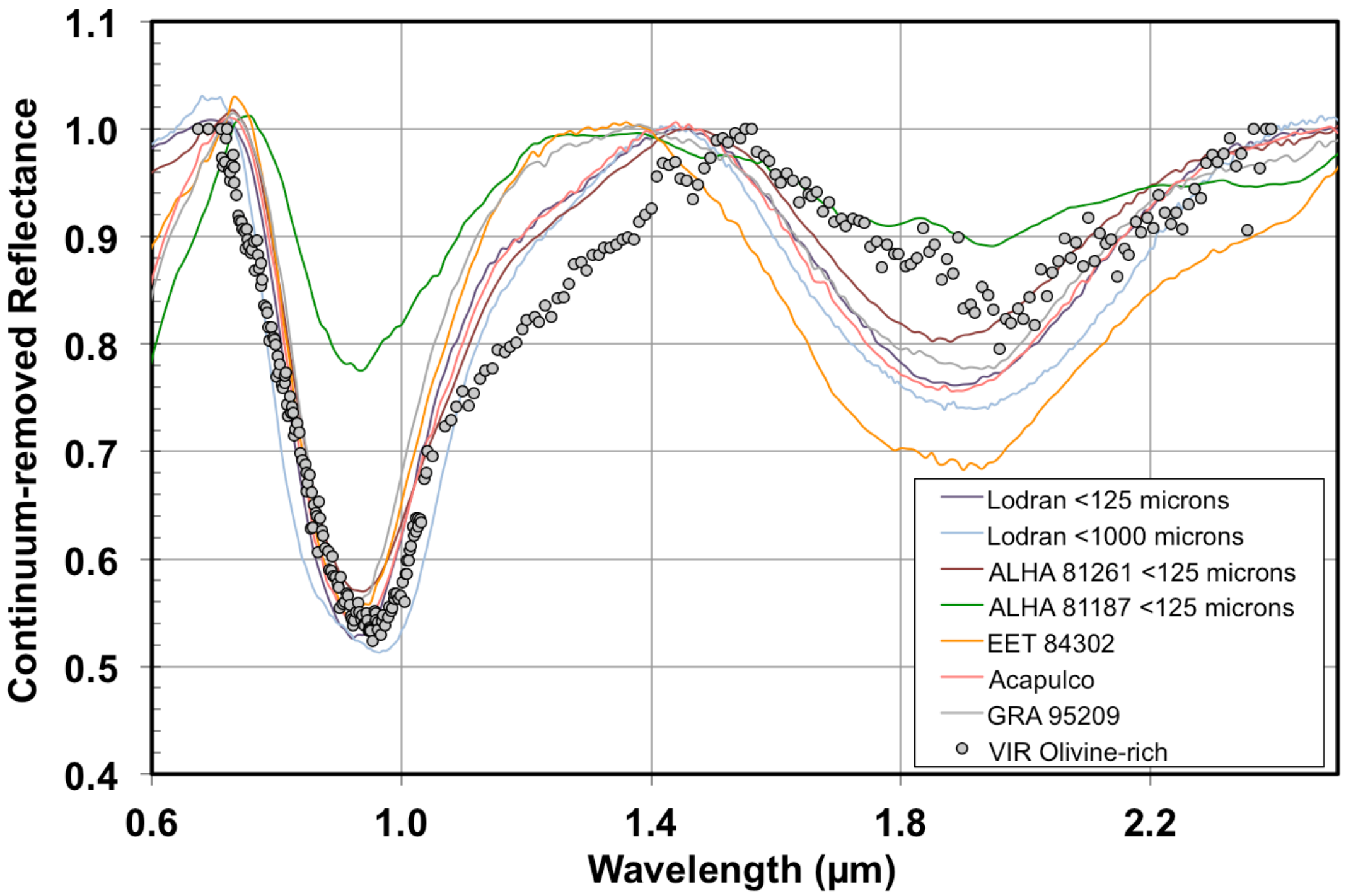

Fig. 3b

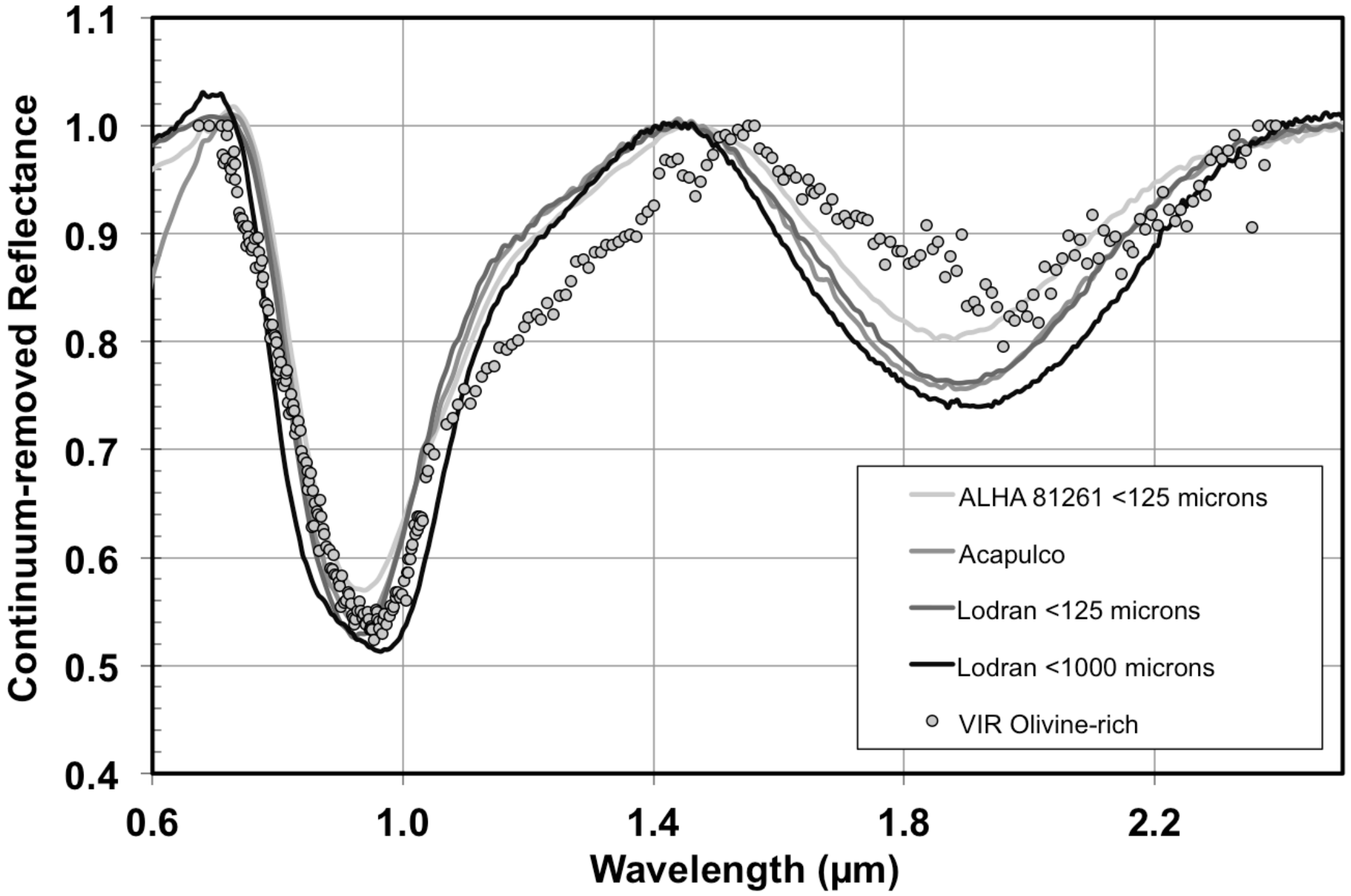

Fig. 4a

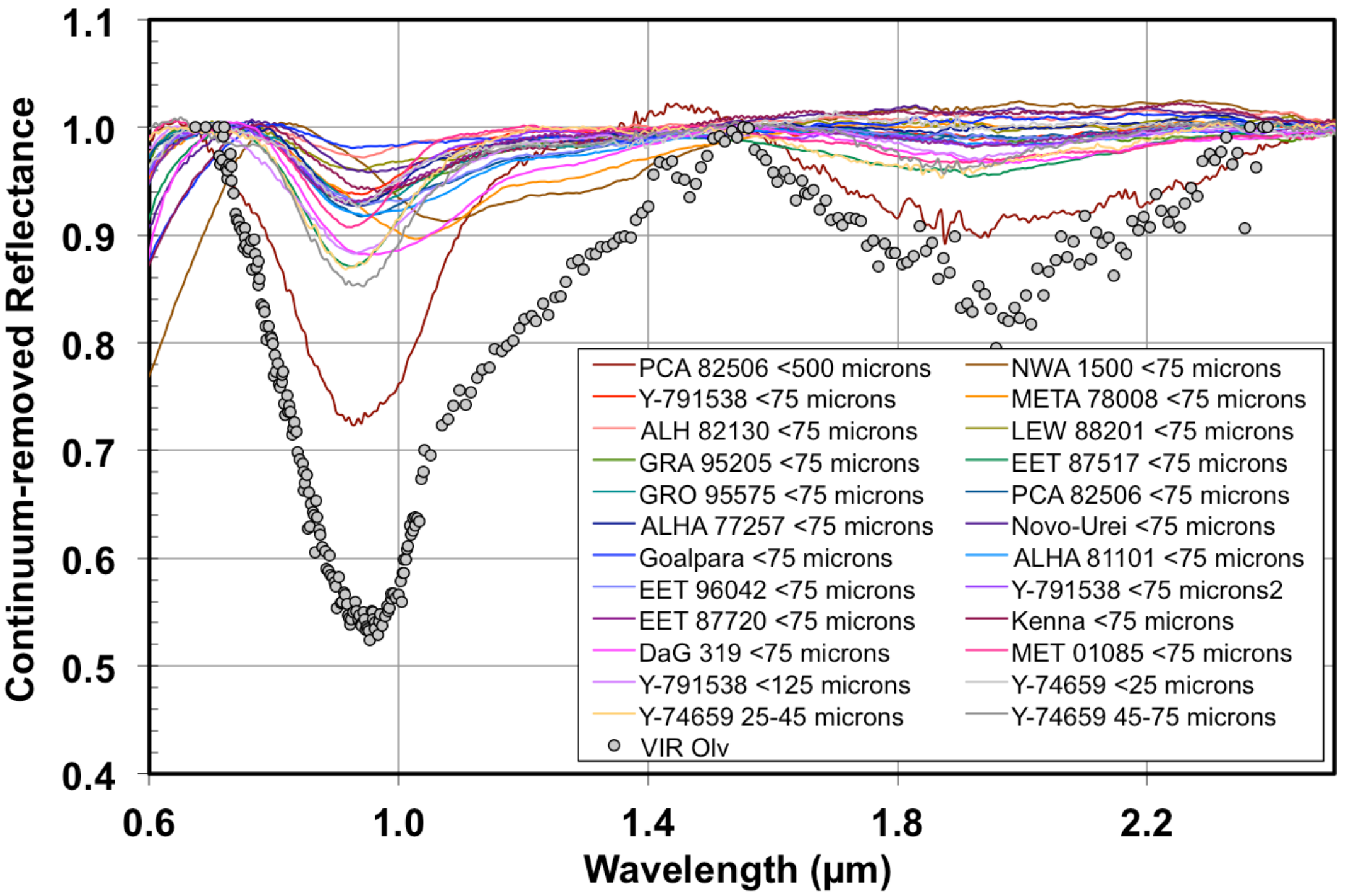

1.1
1.0
0.9
0.8
0.7
0.6
0.5
0.4
Continuum-removed Reflectance
0.6
1.0
1.4
1.8
2.2
Wavelength (μm)
PCA 82506 <500 microns
Y-791538 <75 microns
ALH 82130 <75 microns
GRA 95205 <75 microns
GRO 95575 <75 microns
ALHA 77257 <75 microns
Goalpara <75 microns
EET 96042 <75 microns
EET 87720 <75 microns
DaG 319 <75 microns
Y-791538 <125 microns
Y-74659 25-45 microns
VIR Olv
NWA 1500 <75 microns
META 78008 <75 microns
LEW 88201 <75 microns
EET 87517 <75 microns
PCA 82506 <75 microns
Novo-Urei <75 microns
ALHA 81101 <75 microns
Y-791538 <75 microns2
Kenna <75 microns
MET 01085 <75 microns
Y-74659 <25 microns
Y-74659 45-75 microns

Fig. 4b

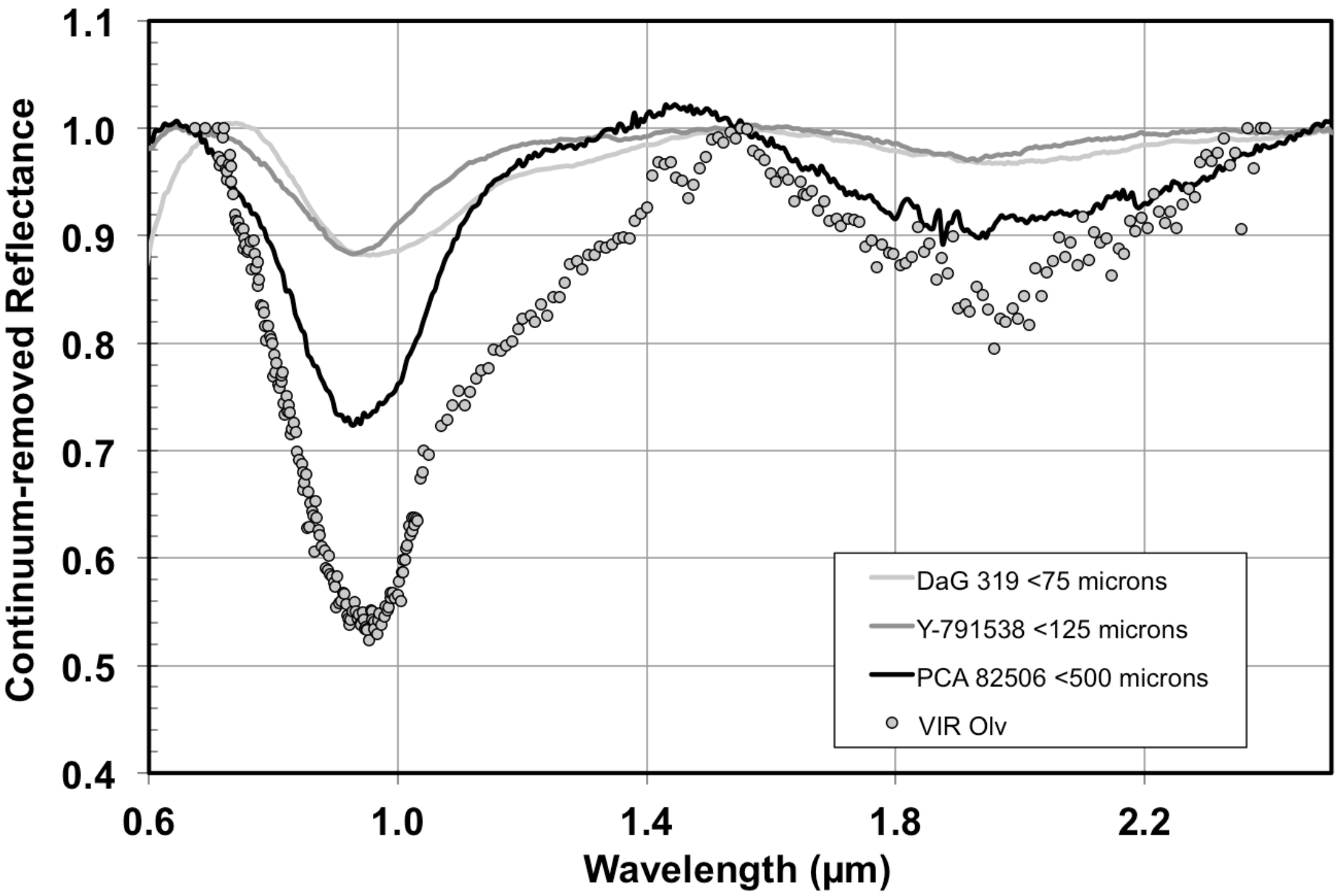

Fig. 5a

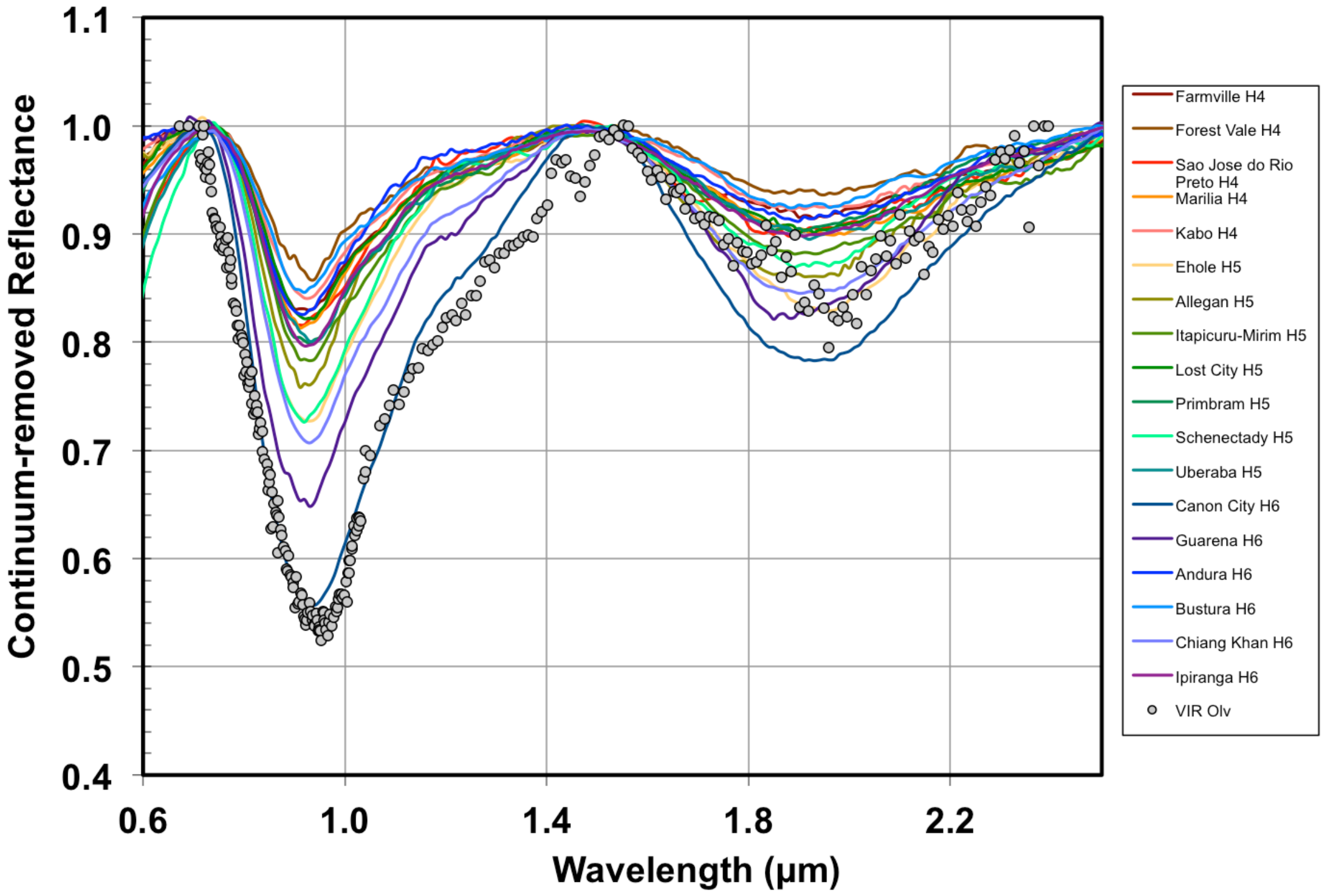

Fig. 5b

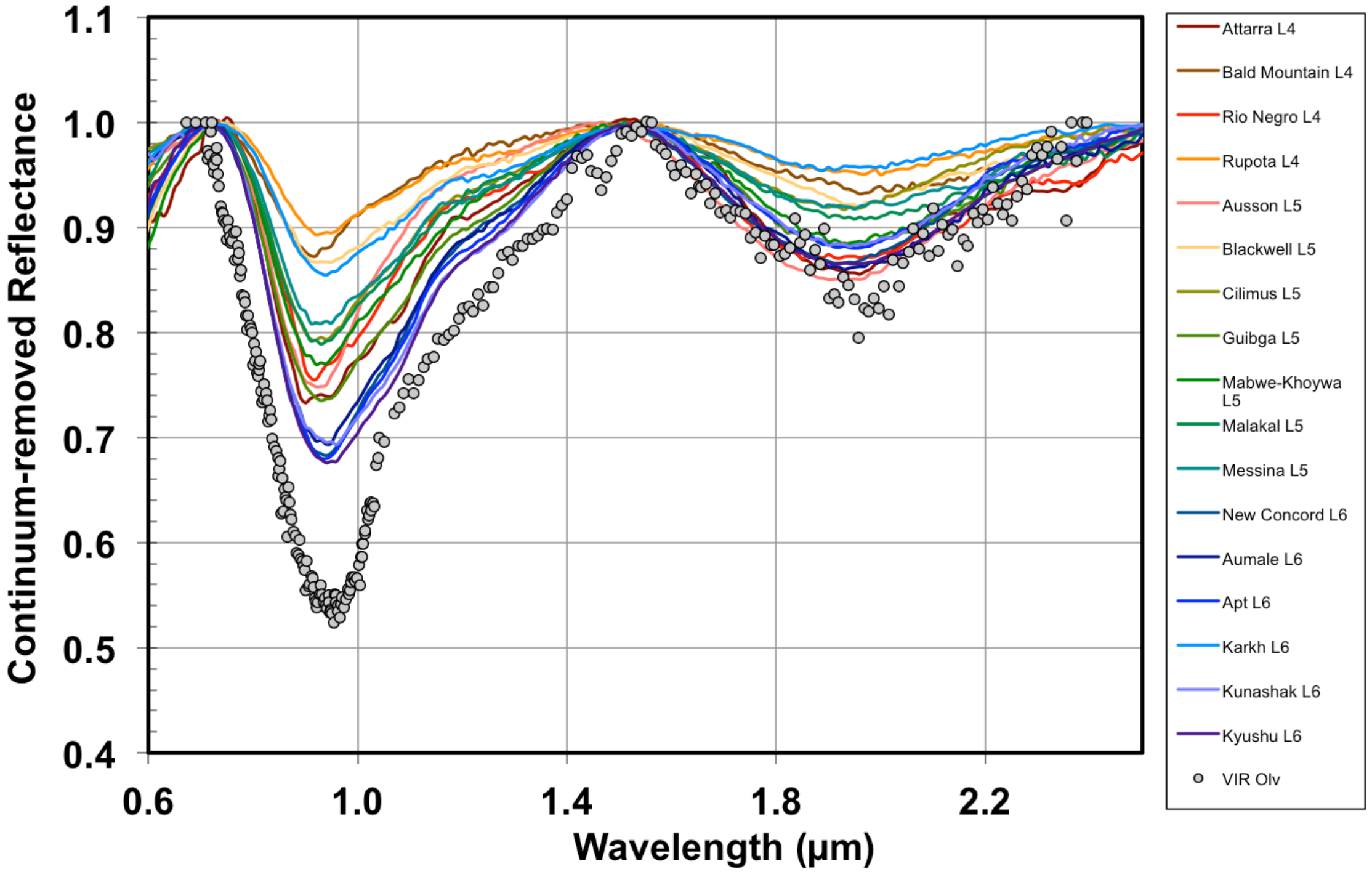

Fig. 5c

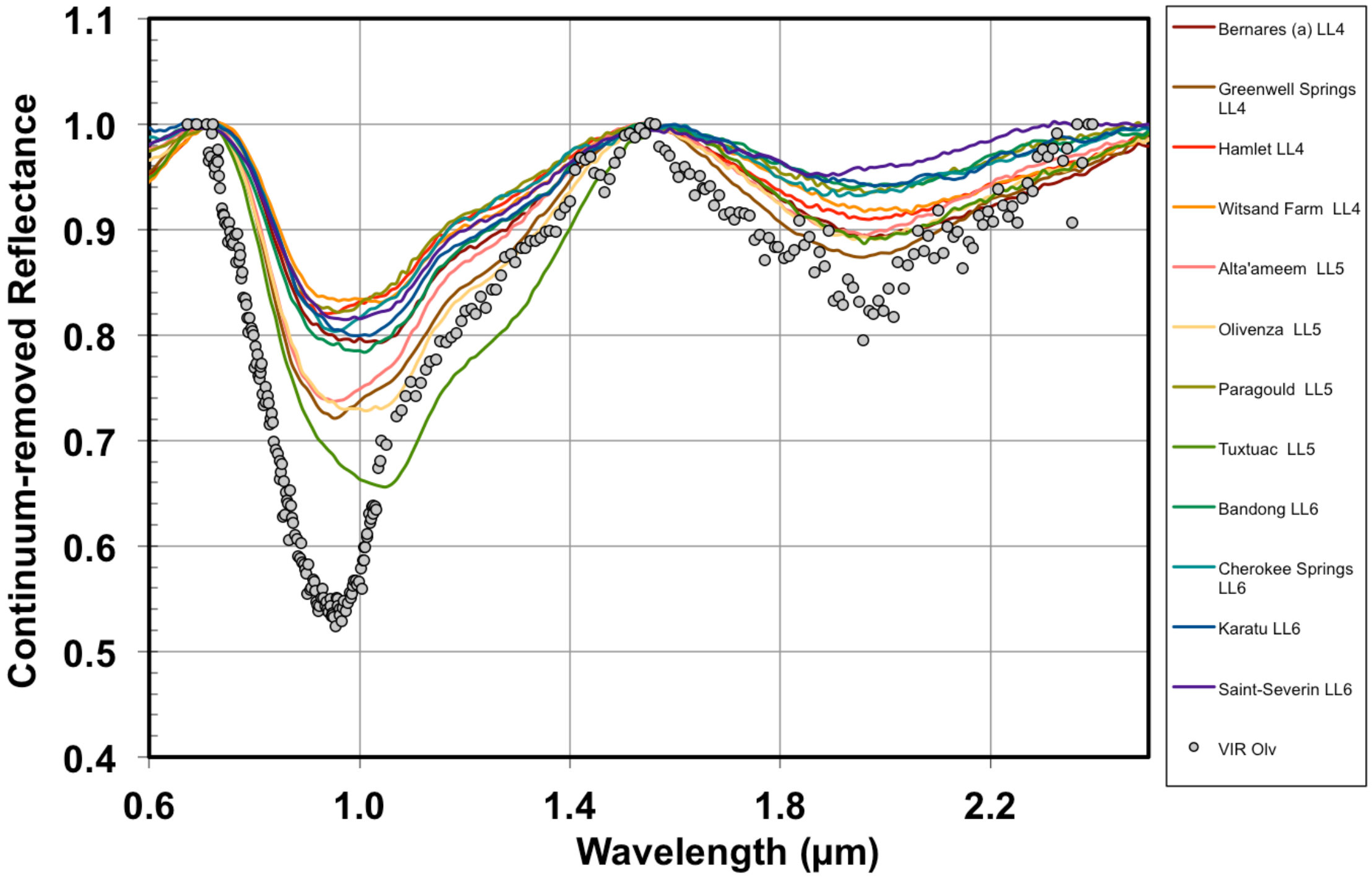

Fig. 5d

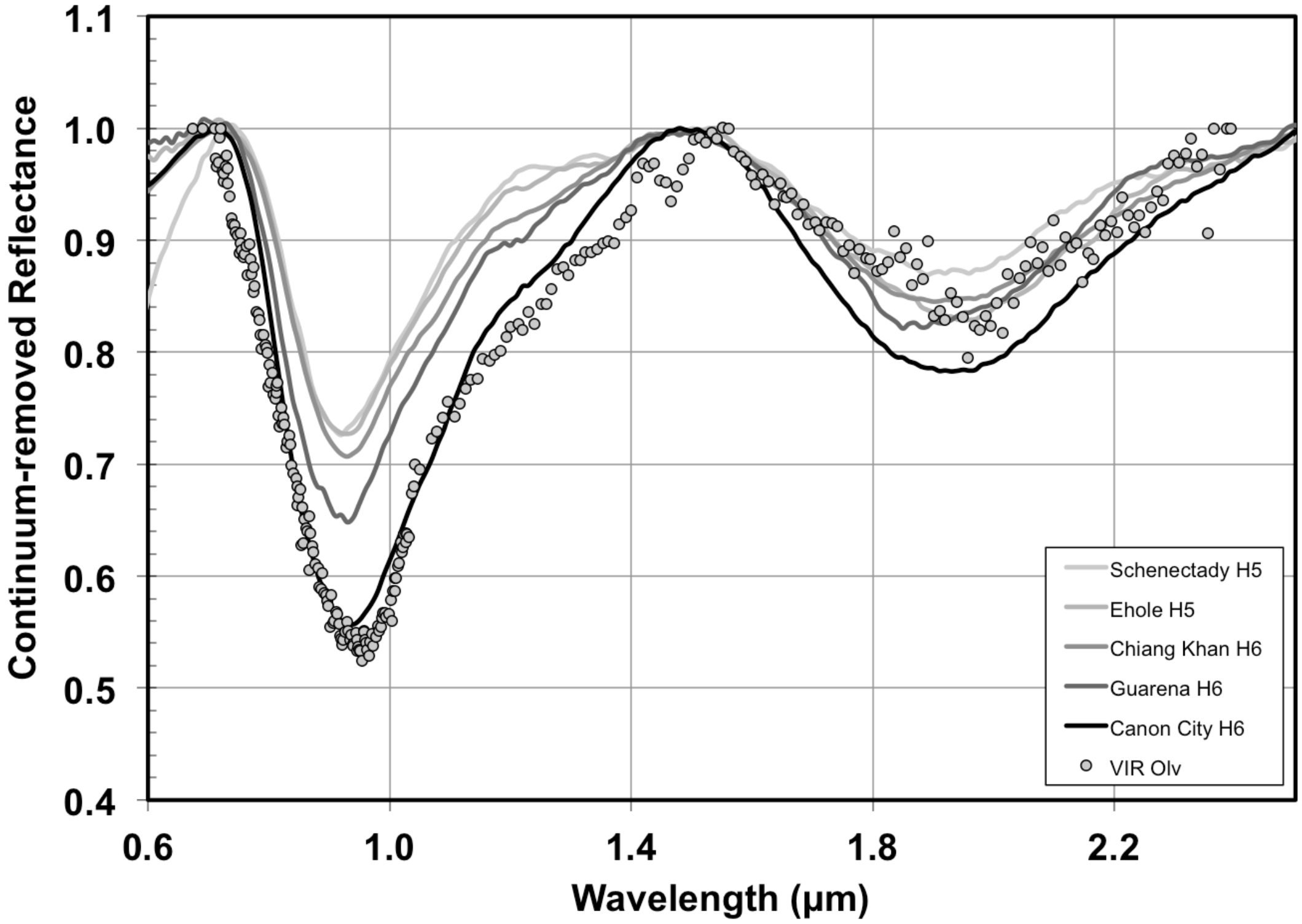

Fig. 5e

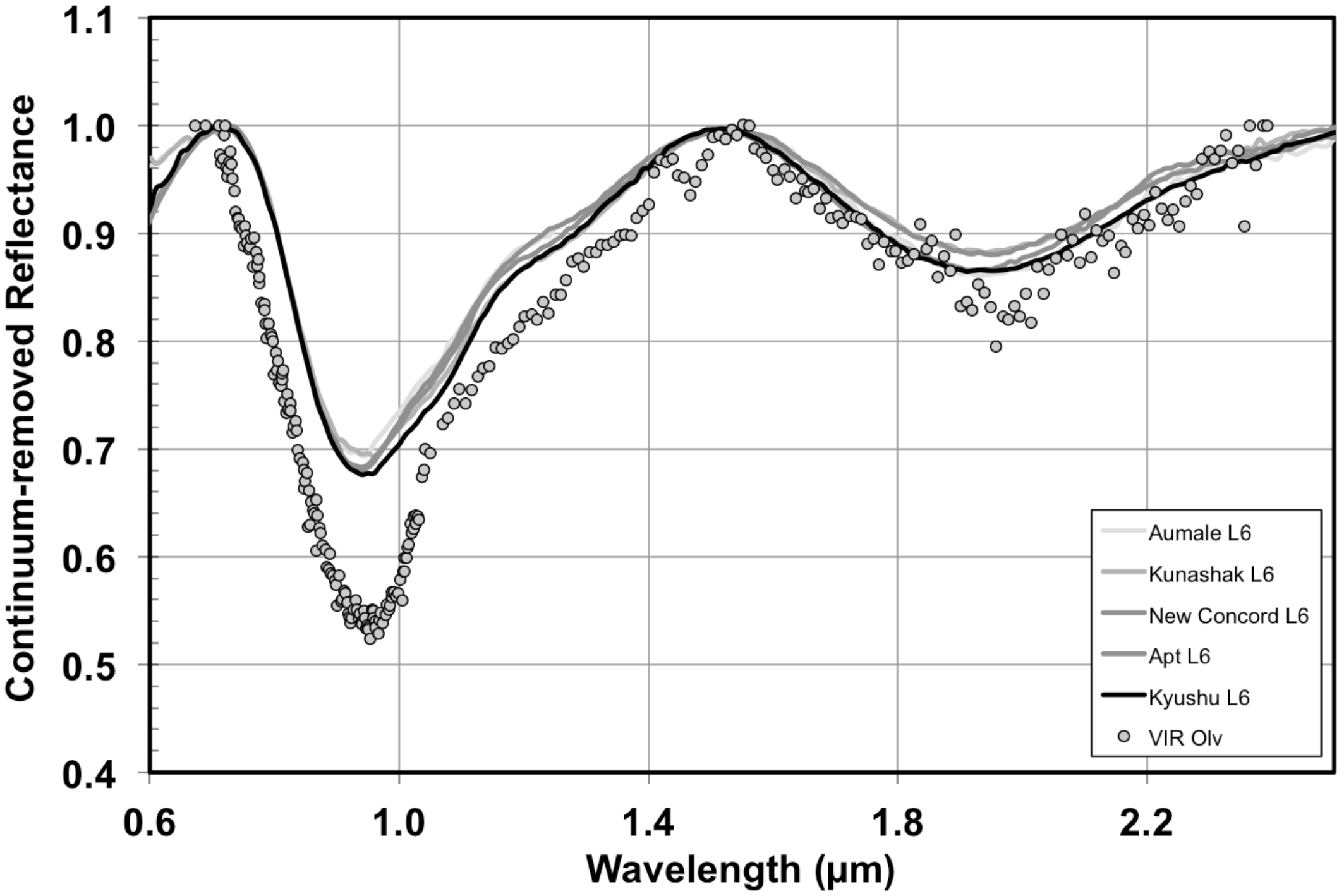

Fig. 5f

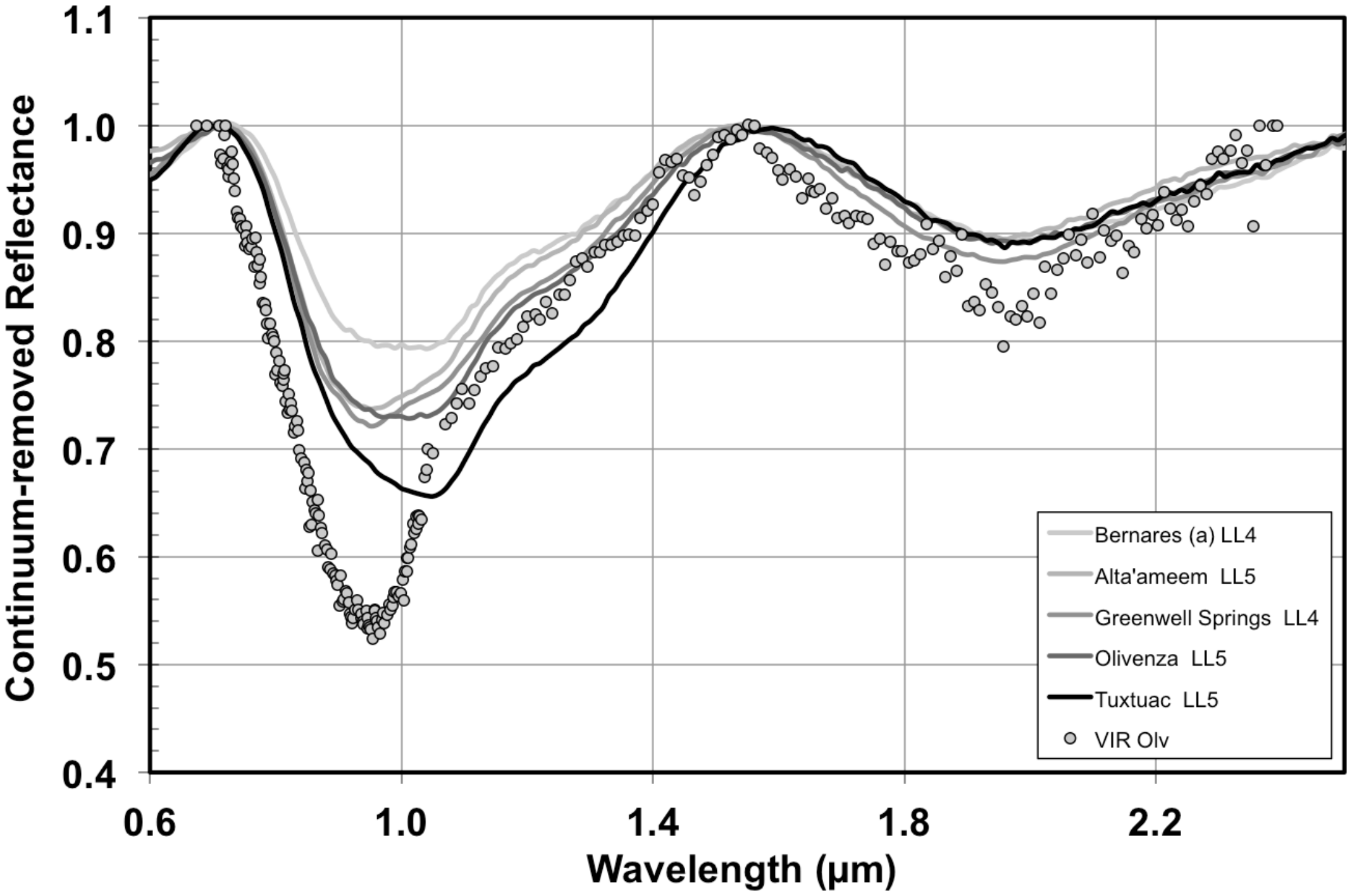

Fig. 6

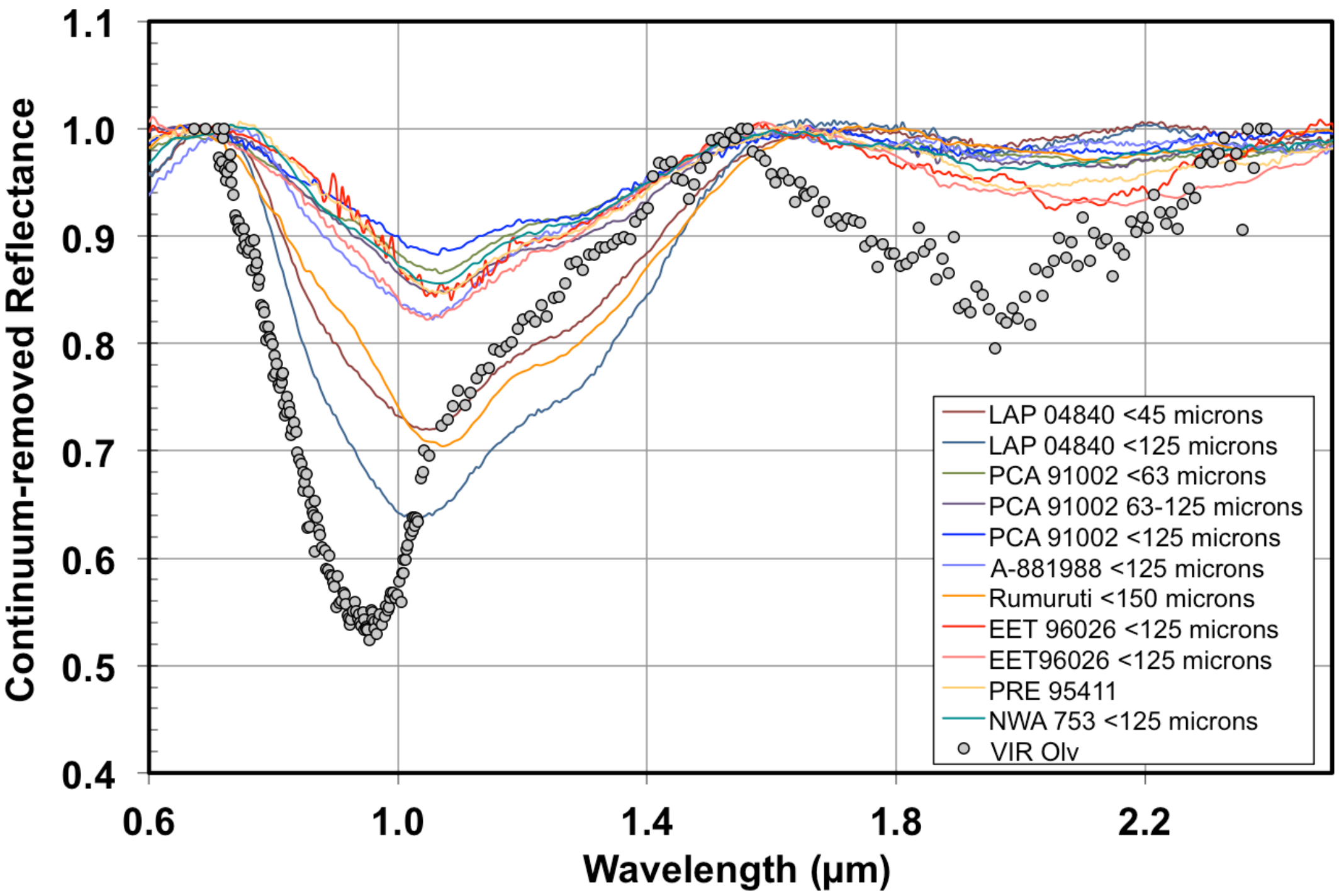

1.1
1.0
0.9
0.8
0.7
0.6
0.5
0.4
Continuum-removed Reflectance
0.6
1.0
1.4
1.8
2.2
Wavelength (µm)
LAP 04840 <45 microns
LAP 04840 <125 microns
PCA 91002 <63 microns
PCA 91002 63-125 microns
PCA 91002 <125 microns
A-881988 <125 microns
Rumuruti <150 microns
EET 96026 <125 microns
EET96026 <125 microns
PRE 95411
NWA 753 <125 microns
VIR Olv

Fig. 7

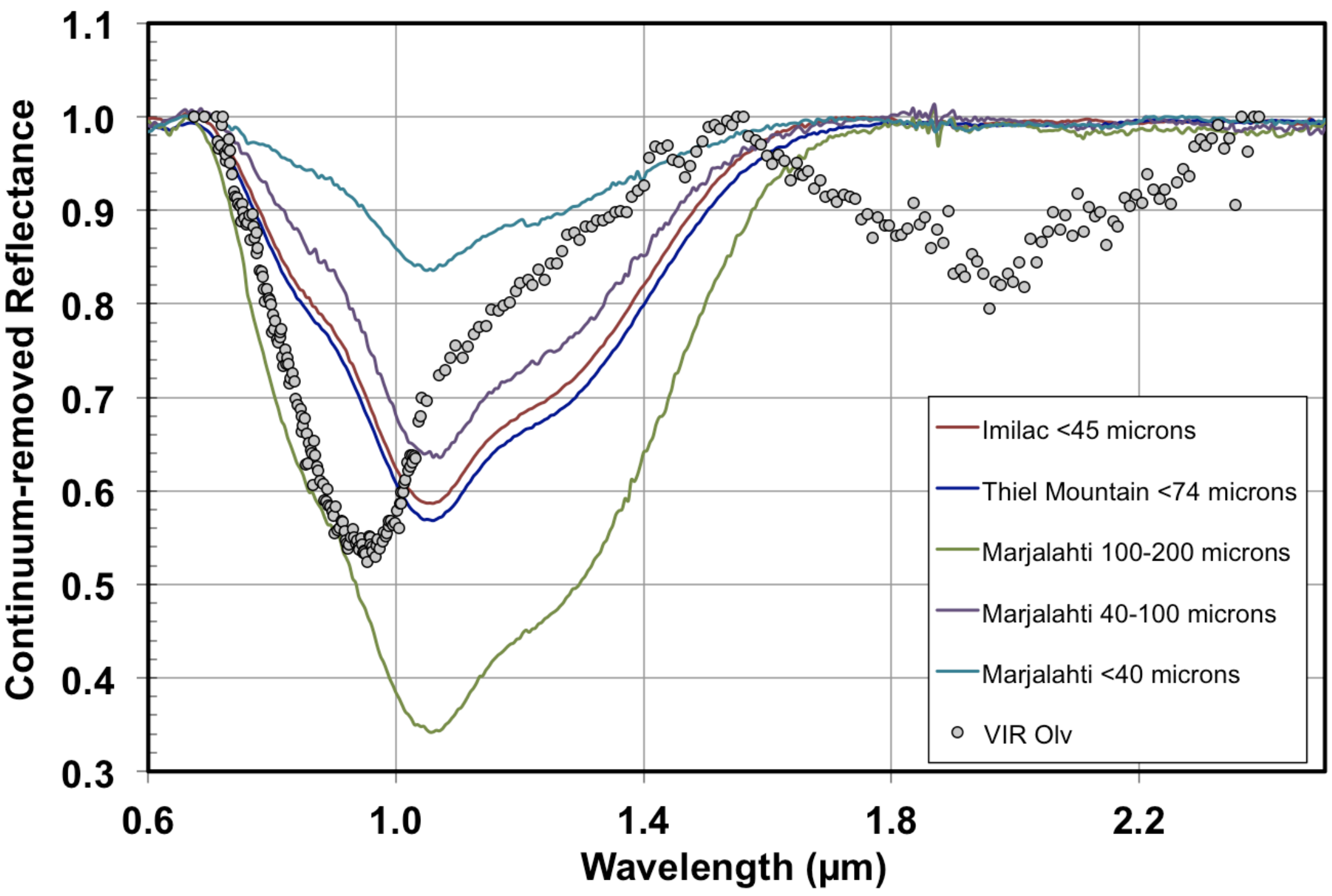

Fig. 8

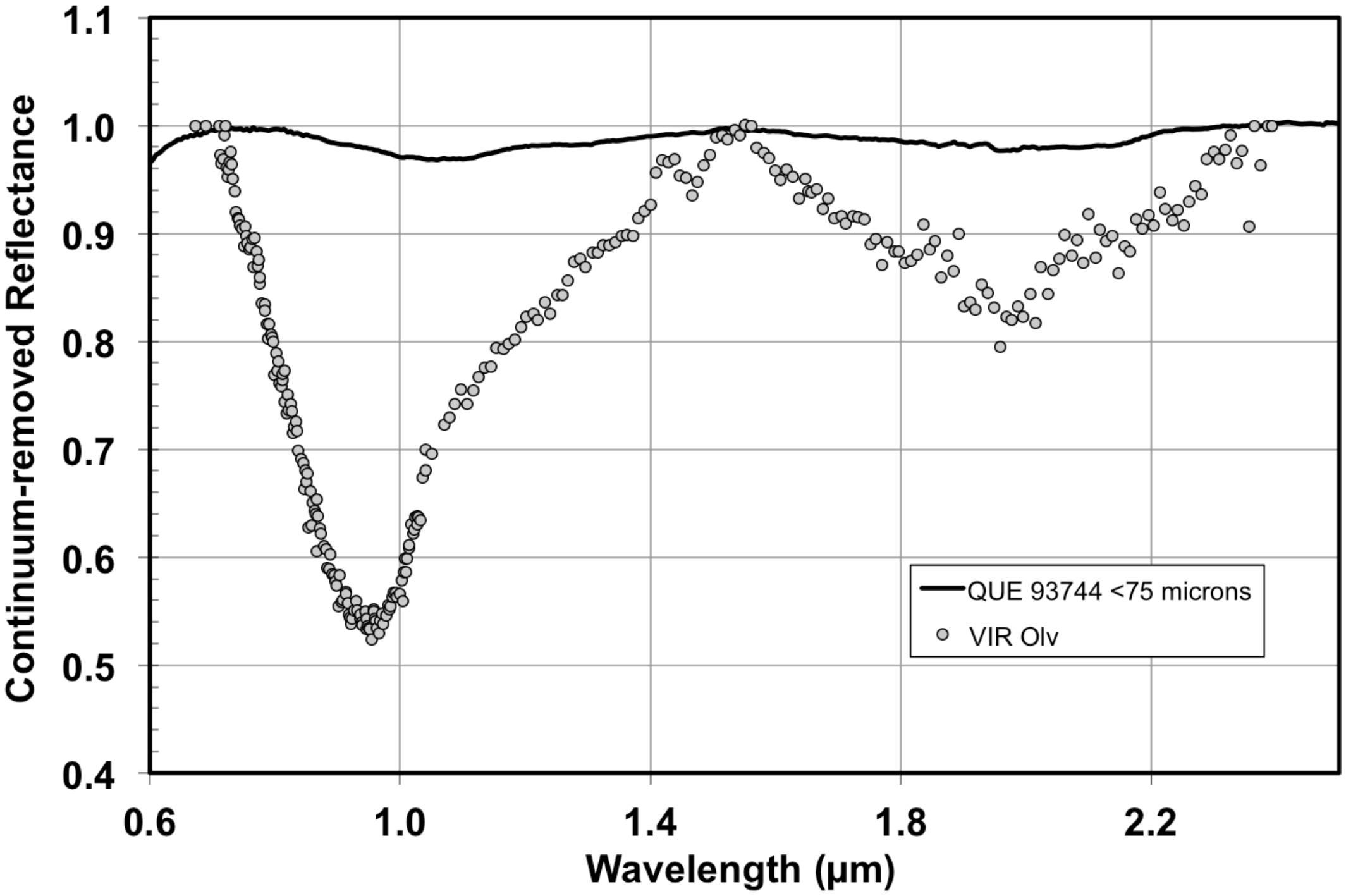

Fig. 9

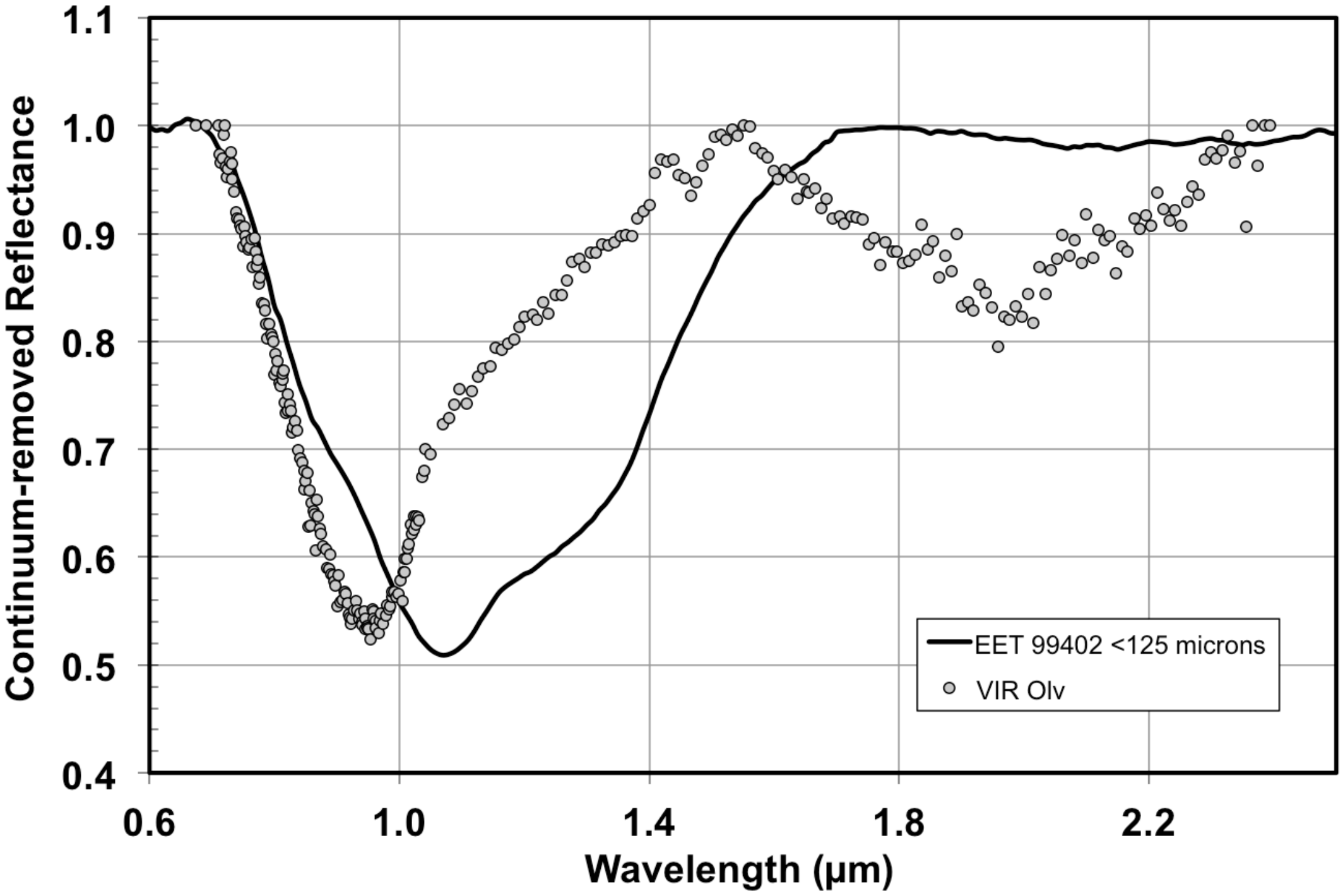

Fig. 10a

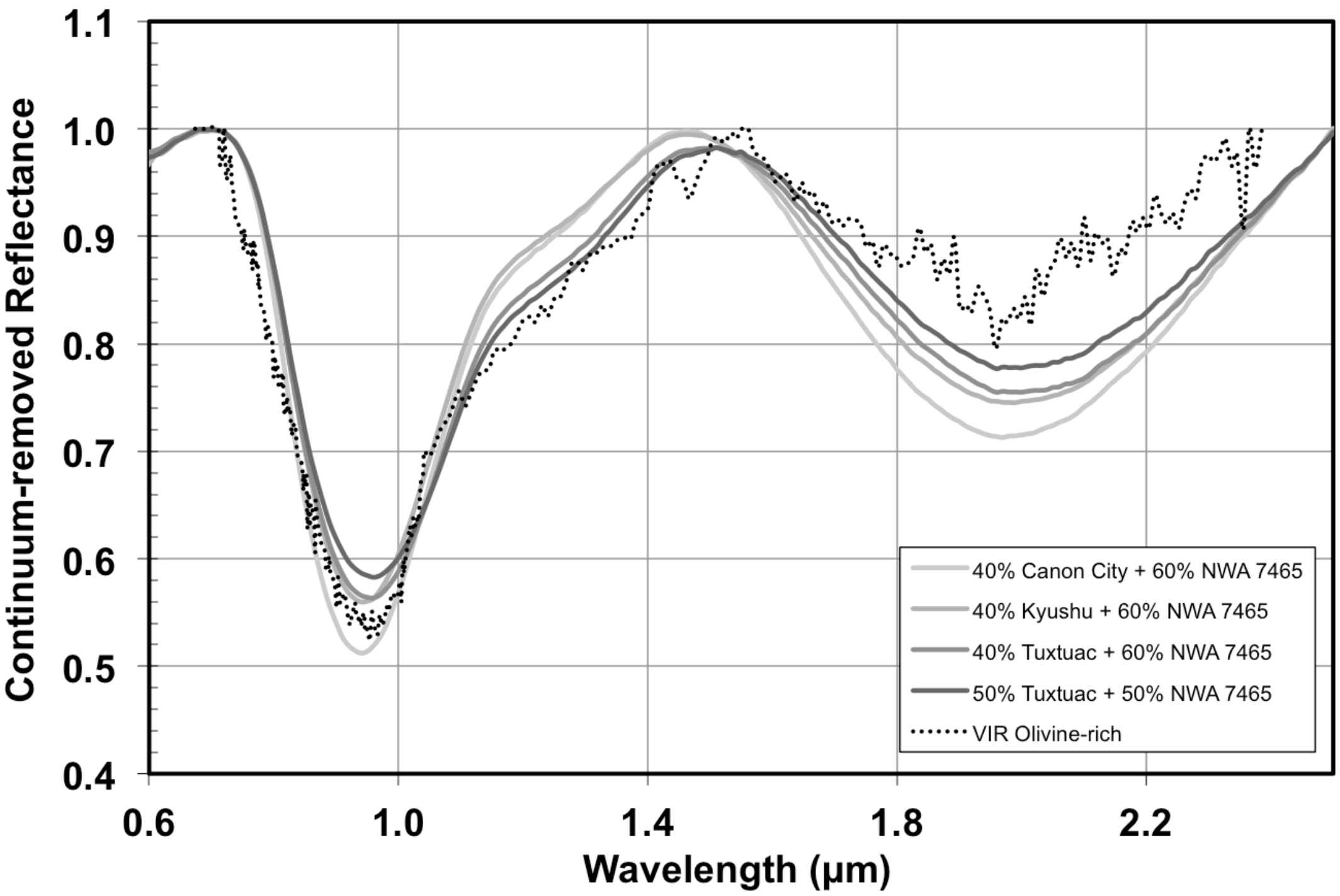

Fig. 10b

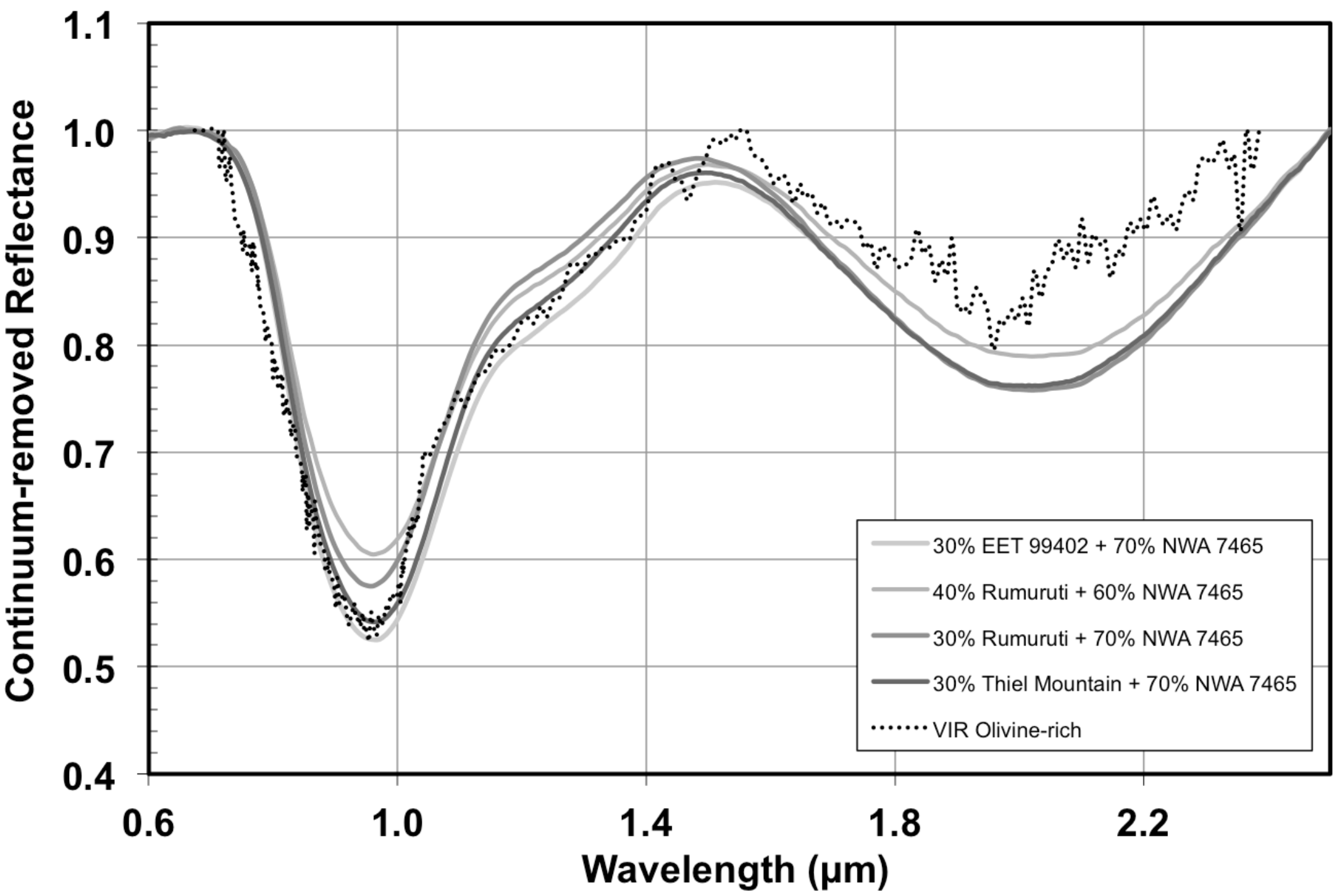

Fig. 11a

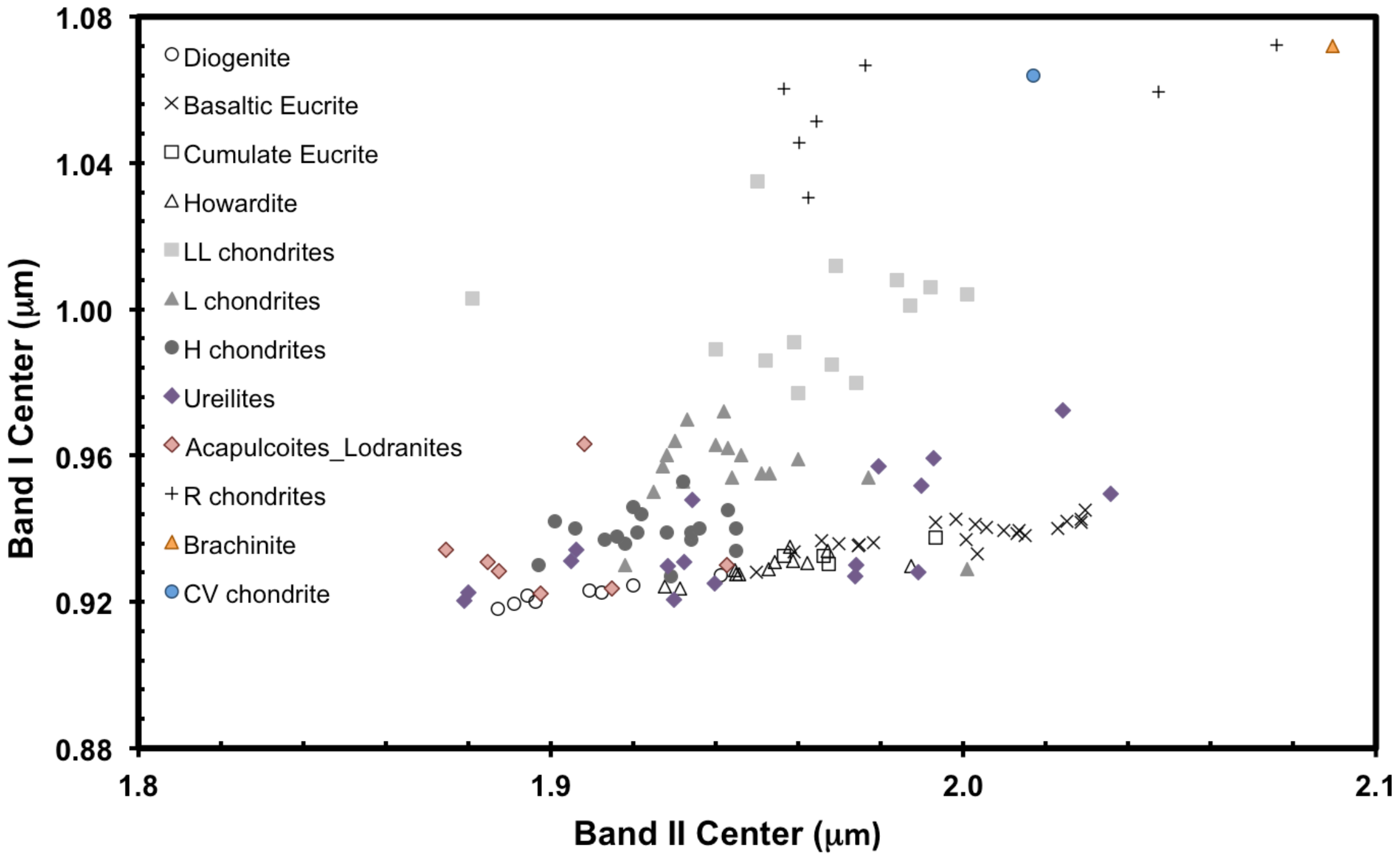

Fig. 11b

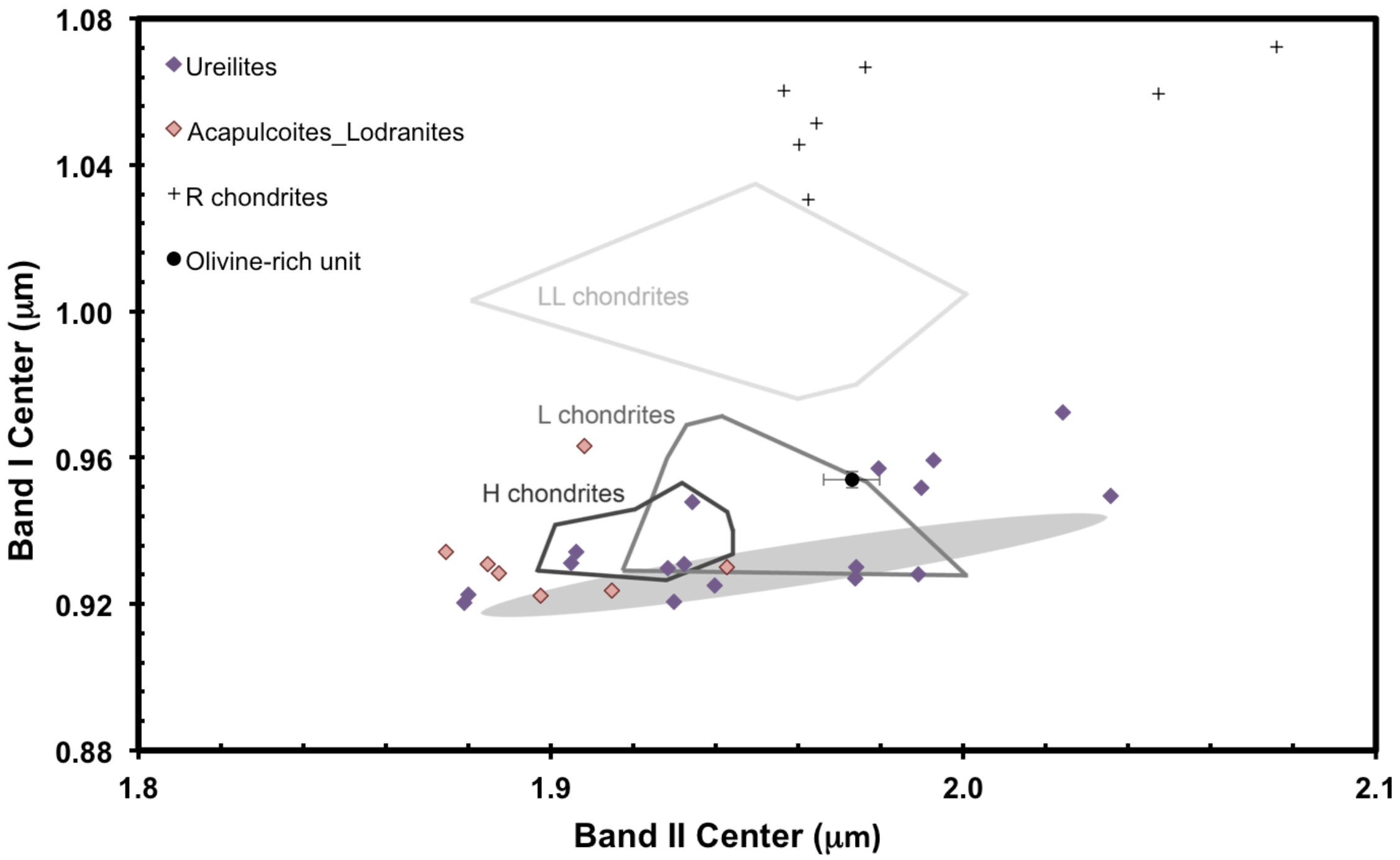

Fig. 12

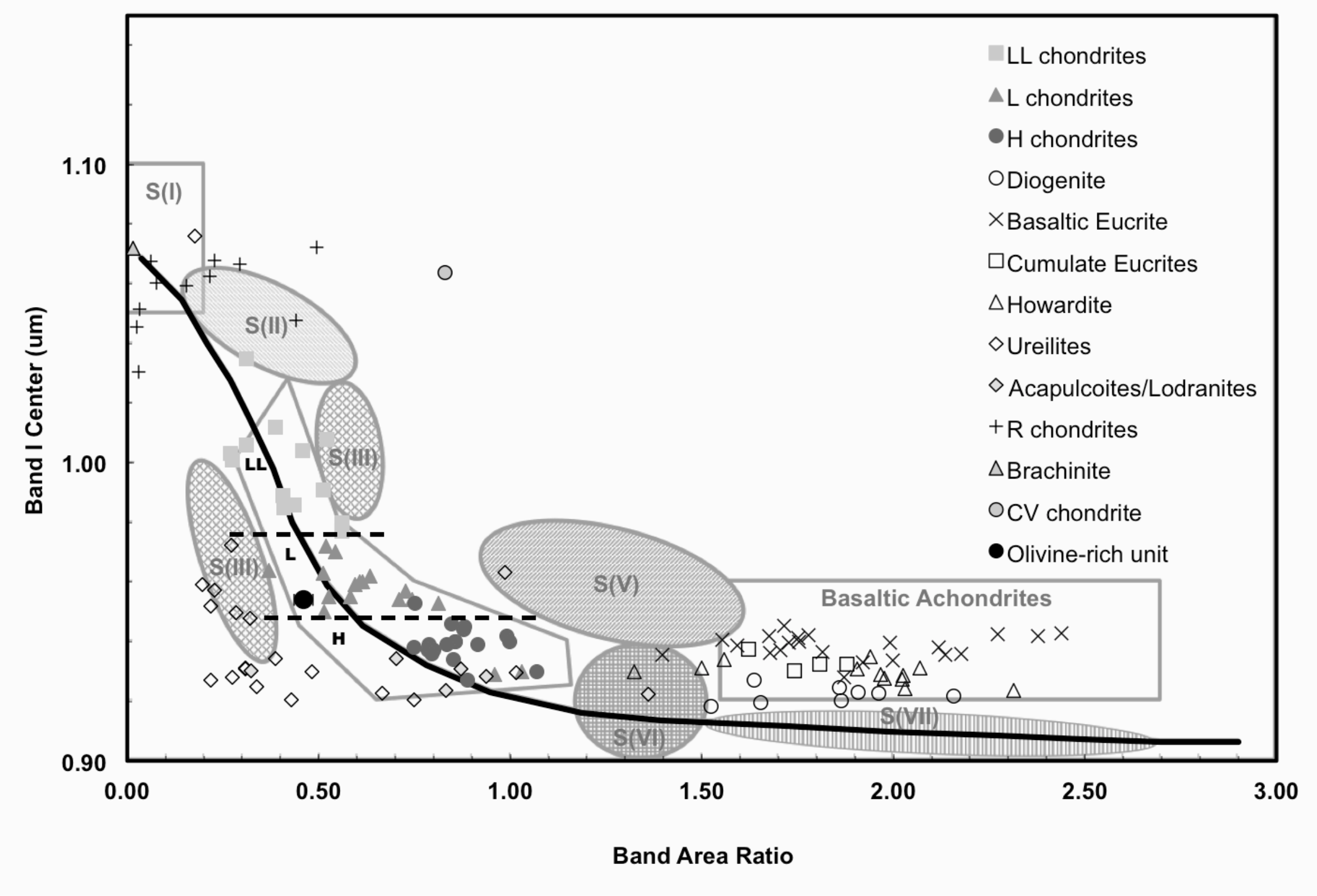